\documentclass[aps,pra,superscriptaddress,showpacs,tightenlines,twocolumn]{revtex4-1}

\usepackage{amsmath}
\usepackage{changepage}
\usepackage{epsfig}
\usepackage{epstopdf}  
\usepackage[colorlinks,citecolor=blue,linkcolor=blue]{hyperref}
\usepackage{lipsum}
\usepackage{amssymb}
\usepackage{mathtools}
\usepackage{subfigure}
\usepackage{multirow}
\usepackage[table]{xcolor}
\usepackage{color}
\usepackage{graphicx}

\begin{document}

\title{Optimal deterministic remote state preparation via a non-maximally entangled channel without additional quantum resources}
\author{Xuanxuan Xin}
\author{Shiwen He}
\author{Yongxing Li}
\author{Chong Li}
\affiliation{School of Physics, Dalian University of Technology, Dalian 116024,China}
\begin{abstract}
    In this paper, we reinvestigate remote state preparation by using the prepared non-maximally entangled channel. An innovative remote state preparation protocol is developed for deterministically preparing information encoded in quantum states without additional consumption of quantum resources. We have increased the success probability of preparing a d-dimensional quantum state to 1 via a non-maximally entangled quantum channel. A feasibly experimental scheme is also designed to realize the above-mentioned deterministic scheme. This work provides a valuable method to address decoherence and environmental noises in the practicalization of quantum communication tasks.
\end{abstract}

\maketitle
\section{Introduction}
The quantum network, the backbone of quantum computing architectures, is composed of spatially separated nodes storing quantum information in quantum bits and connected by quantum channels \cite{PhysRevA.29.1419, PhysRevLett.78.3221, ritter2012elementary, simon2017towards, PhysRevLett.120.030501}. The exchange of quantum information between different nodes is accomplished via quantum channels. Remote state preparation (RSP) can be considered as transferring known quantum information between two different quantum nodes \cite{PhysRevA.62.012313}. Two communicators called Alice and Bob are placed in two different quantum nodes and previously share an entangled quantum state as the quantum channel. In ideal RSP protocols \cite{PhysRevLett.87.077902, pati2000minimum, nguyen2011remote}, Alice deterministically prepares a known quantum state for Bob through a maximally entangled quantum channel. Due to decoherence and environmental noise, the maximally entangled channel employed in RSP schemes inevitably degenerates into an undesired non-maximally entangled one \cite{PhysRevA.52.R2493, PhysRevLett.70.1187, PhysRevLett.81.2594, RevModPhys.76.1267, PhysRevA.72.012315}, the success probability of the RSP is degraded from 100\% correspondingly.

Entanglement is indispensable to the application of quantum technologies, like scalable quantum calculation \cite{kimble2008quantum,pirandola2016physics,wehner2018quantum} and quantum communication assignments \cite{PhysRevLett.69.2881,PhysRevLett.70.1895,PhysRevLett.85.441,RevModPhys.81.1301,pirandola2015advances}. It is lighthearted to manipulate with the current experimental techniques whereas it is still challenging to entangle different long-distance quantum systems strongly \cite{dakic2012quantum,PhysRevLett.106.130506,yao2012observation,huang2011experimental}. Compared to the demanding maximally entangled channel employed in the standard RSP protocol, it is more readily to prepare generally entangled states as quantum resources for transferring quantum information \cite{wagenknecht2010experimental,zhang2011preparation,yin2020entanglement}. However, employing a non-maximally entangled channel is less efficient than utilizing a maximally entangled one to transfer quantum information conventionally \cite{PhysRevA.62.012313, PhysRevLett.87.077902, PhysRevA.63.014302}. Therefore, one of the problems of fundamental and practical significance is to engineer deterministic RSP (DRSP) protocols with excellent robustness against decoherence and environmental noises. The mainstream solution currently is to increase the success probability of RSP by enhancing the entanglement of quantum channels. There exist sorts of practices to enhance entanglement, such as entanglement purification \cite{PhysRevLett.76.722,PhysRevLett.77.2818,PhysRevA.54.3824,PhysRevLett.97.180501,PhysRevLett.110.260503,PhysRevLett.126.010503}, quantum catalysis \cite{jonathan1999entanglement,PhysRevA.64.042314,PhysRevA.67.060302,PhysRevA.79.054302} and local filtering operations \cite{PhysRevLett.74.2619,PhysRevA.54.2685,PhysRevLett.96.150501,PhysRevLett.100.090403,PhysRevA.86.052115}. Entanglement purification protocols achieve the above objective by locally manipulating multiple copies of harsh entangled states to produce fewer copies with increased fidelity \cite{PhysRevLett.127.040502}. The method with the advantage of quantum catalysts has high communication efficiency without ancillary entanglement consumed or degraded \cite{PhysRevLett.127.080502}. However, more entanglement resources are requisite beforehand in both approaches, which increases the difficulty of experiments. Local filtering operations on entangled quantum channels may activate their desirable features but additionally have higher experimental complexity \cite{PhysRevResearch.3.023045}. Briefly, boosting the degree of entanglement of quantum channels requires tedious operations and more consumption of quantum resources. It is still challenging experimentally so far to entangle diverse systems strongly. Therefore, it is worth considering whether there exist other ingenious approaches instead of improvement on the entanglement strength of quantum channels to increase the success probability of RSP.

Herein, we propose an optimal DRSP protocol where any given quantum state is accessible to be prepared deterministically via a generally entangled quantum channel. The probability of success has increased from $2|\alpha|^{2} (|\alpha|^{2}\leq\frac{1}{2})$ to one without increased quantum resources, and it is scarcely affected by the entanglement of quantum channels in this well-designed scheme. Even if the utilized entangled channel is degraded by environmental noises, the probability of success stays constant as long as this entanglement is present. It is unnecessary to increase consumption of quantum resources to improve the quality of quantum channels beforehand, which minimizes the expenditure of costs and lessens the experimental complexities significantly. For specific operations, measurement procedures involved in this RSP programme are scheduled in the last step for better simplicity, rather than interspersed in the scenario as before. And the easier-to-operate projection measurement (PM) under the simple set of orthogonal vectors is adopted, not the positive operator-valued measurement inapproachable experimentally \cite{PhysRevA.69.022310, nguyen2011remote, PhysRevA.65.022316, PhysRevA.98.042329}. This DRSP protocol provides constructive implications for the further development of quantum communication technologies.

\section{Deterministic remote preparation of a d-dimensional quantum state via a generally entangled quantum channel}
\begin{figure}[htbp]
    \includegraphics[width=0.4\paperwidth]{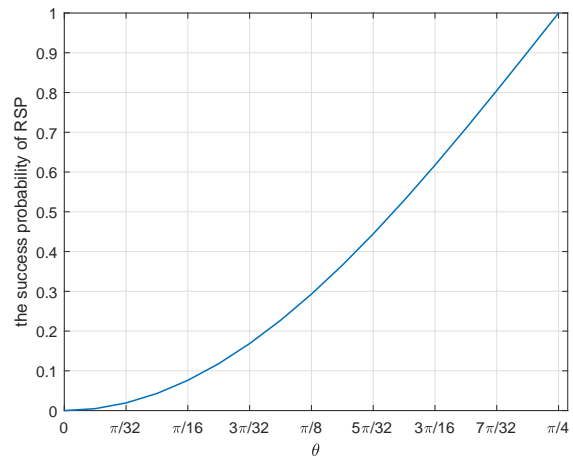}
    \caption{Schematic representation of the success probability of RSP as a function of the entanglement coefficient $\theta$ in conventional RSP schemes. It is  presented that the probability of success can reach 1 only when $\theta=\frac{\pi}{4}$, i.e., the quantum channel is a maximally entangled quantum state ($|\alpha|=|\beta|=\frac{1}{\sqrt{2}}$).}\label{Fig.1}

\end{figure}

\emph{Conventional RSP protocols} - Let us recall the standard RSP protocol briefly \cite{PhysRevA.62.012313,PhysRevLett.87.077902,pati2000minimum}. Two spatially separated parties, Alice and Bob, previously share a maximally entangled quantum channel $|\varphi\rangle_{AB}=\frac{1}{\sqrt{2}}(|00\rangle+|00\rangle)_{AB}$. Alice possesses one of the entangled pairs qubit A and Bob possesses the other qubit B. The goal of Alice is to prepare information encoded on a qubit for Bob, using local operations and classical communication (LOCC). The qubit state that Alice aims to prepare for Bob is generalised as $|\varphi\rangle=x_{0}|0\rangle+x_{1}|1\rangle$, where $x_{0}$ and $x_{1}$ are complex numbers. Alice initially measures her member of the preshared pair under the basis associated with the prepared information and sends the outcome to Bob via a classical-communication channel. On receiving the measurement outcome, Bob performs corresponding unitary transformations on his member of the shared pair to recover the unknown qubit. For RSP to invariably succeed via a maximally entangled channel, Nguyen \emph{et. al} modulated the standard RSP protocol slightly \cite{nguyen2011remote}: Alice extra adds an auxiliary qubit named C and performs a C-NOT gate \cite{PhysRevA.60.2777, PhysRevLett.101.250501, PhysRevLett.100.160502} on qubits A and C. She respectively measures these two qubits on two various bases concerning the prepared information and sends outcomes to Bob. Bob thereby performs relevant unitary operations on qubit B. The weakness of the above RSP schemes is that they are only applicable to preparing 2, 4, or 8-dimensional quantum states \cite{PhysRevA.65.022316}. Nguyen \emph{et. al} constructed another DRSP protocol afterward for any dimension employing positive operator-valued measurement \cite{PhysRevA.98.042329}. Regrettably, these quantum communication programs are accomplished via pure maximally entangled channels. Realistic quantum channels are noisy inevitably on account of decoherence and environmental factors. While the quantum channel is not a maximally entangled state but a partially entangled one $|\varphi\rangle_{AB}=(\alpha|00\rangle+\beta|11\rangle)_{AB}$ ($|\alpha|^{2}+|\beta|^{2}=1$, $|\alpha|\leq|\beta|$), the success probability is decayed to $2|\alpha|^{2} (|\alpha|^{2}\leq\frac{1}{2})$ consequently in previous RSP schemes (further detailed derivation can be found in Appendix \ref{app1}). Assuming $|\alpha|=sin\theta$, $|\beta|=cos\theta (0 \le \theta \le \frac{\pi}{4})$, then the success probability of RSP can be represented by $2(sin\theta)^2$. Fig. \ref{Fig.1} illustrates the variation of the success probability as a function of $\theta$. Undoubtedly, the closer $\theta$ is to $\frac{\pi}{4}$, the higher the success probability is. Only when $\theta=\frac{\pi}{4}$, i.e., employing a maximally entangled channel, does the probability of success reach 1 according to traditional RSP protocols. To that end, it is of absolute necessity to design a refreshed protocol in which RSP invariably succeeds even if the quantum channel is a non-maximally entangled one ($|\alpha| \neq |\beta|$).

\emph{The optimal DRSP protocol} - To deterministically preparing quantum information via a non-maximally entangled channel, we have designed a generalized DRSP scheme where quantum information encoded in the quantum state can be prepared faithfully even if the quantum channel is a non-maximally entangled state. The success probability of RSP is increased from $2|\alpha|^{2} (|\alpha|^{2}\leq\frac{1}{2})$ to 1 by means of local opearations and simpler measurement approaches. The following is a detailed description of the optimal DRSP scheme proposed in this investigation. Assuming that Alice assists Bob in the preparation of a d-dimensional quantum state whose mathematical formalism can be generalized as follows, 
\begin{eqnarray}
    \begin{aligned}
        |\varphi\rangle=&\sum_{n=0}^{d-1}x_{n}|n\rangle\\
        =&\sum_{n=0}^{d-1} |x_{n}|e^{i\theta_{n}}|n\rangle \label{equation 1},
    \end{aligned} 
\end{eqnarray}
where $x_{n}$ are complex numbers satisfying the normalization condition $\sum_{n=0}^{d-1}|x_{n}|^{2}=1$. Qudits A and B are entangled and possessed by Alice and Bob severally. The quantum channel is the entangled quantum state $|\varphi\rangle_{AB}$ represented by the following form, 
\begin{eqnarray}
    \begin{aligned}
        |\varphi\rangle_{AB}=\sum_{m,n=0}^{d-1} \lambda_{mn}|mn\rangle_{AB},
    \end{aligned} 
\end{eqnarray}
where complex numbers $\lambda_{mn}$ satisfy orthogonal normalization $\sum_{m,n=0}^{d-1}|\lambda_{mn}|^{2}=1$. Only if all $|\lambda_{mn}|$ are equal to the same value, $|\varphi\rangle_{AB}$ is a maximally entangled quantum state. As in a DRSP scheme employing a maximally entangled channel \cite{nguyen2011remote}, Alice introduces an auxiliary qudit C set in the initial state $|0\rangle_{C}$ and employs a C-NOT gate on qudits A and C, which entangles these three particles and renders two separable quantum states $|\varphi\rangle_{AB}$ and $|0\rangle_{C}$ into an hyperentangled quantum state
\begin{eqnarray}
    \begin{aligned}
        |\varphi_{0}\rangle_{ABC}=&C_{AC}^{(m,j)}(|\varphi\rangle_{AB}\otimes|0\rangle_{C})\\
        =&\sum_{m,n=0}^{d-1}\lambda_{mn}|mnk_{m}\rangle_{ABC},
    \end{aligned} 
\end{eqnarray}
the mathematical form of a C-NOT gate can be witten as
\begin{eqnarray}
    \begin{aligned}
        C_{AC}^{(m,j)}=\sum_{m,j=0}^{d-1} |m,j \oplus k_{m}\rangle \langle m,j|,
    \end{aligned} 
\end{eqnarray}
where $k_{m}=0,1,\cdots,d-1$. We hereby assume that $j \oplus k_{m}=k_{m}$ when $j=0$. The following operations are disparate from the traditional RSP schemes. A series of artful operations would be applied to realize DRSP via a non-maximally entangled channel. Firstly,  Alice manipulates qudits A and C unitarily and then transmits qudit A to Bob while the information to be prepared is stored in the hyperentangled quantum state $|\varphi_2\rangle_{ABC}$. Thereby Bob can operate on qudits A and B locally. In the end, Alice and Bob perform PM on qudits C and A under the orthogonal basis \{0,1,...,d-1\} respectively. What follows is an exhaustive elaboration of the above tactics.

\textbf{Step (\uppercase\expandafter{\romannumeral1})} Since Alice knows the information prepared for Bob, she is allowed to perform the unitary transformation associated with the information on qudit A. After that the hyperentangled state is unitarily evolved into
\begin{eqnarray}
    \begin{aligned}
        |\varphi_{1}\rangle_{ABC}=&U^{m}_{A}|\varphi_{0}\rangle_{ABC}\\
                                 =&\sum_{m,n=0}^{d-1} \sum_{s=0}^{d-1} \lambda_{sn}|x_{s}|e^{i(\theta_{s}\pm \theta_{m,s})}|snk_{m}\rangle_{ABC},
    \end{aligned} 
\end{eqnarray}
where
\begin{eqnarray}
    \begin{aligned}
U^{m}_{A}=\sum_{m=0}^{d-1} \sum_{s=0}^{d-1}  |x_{s}|e^{i(\theta_{n}\pm \theta_{m,s})}|s\rangle \langle m|.
\end{aligned} 
\end{eqnarray}

\textbf{Step (\uppercase\expandafter{\romannumeral2})} Alice employs a quantum phase gate (QPG)  on qudits A and C. The function of QPG is that one particle obtains an additional phase regarding the phase information conditional to the other particle state according to the transformation $|i\rangle_{1}|j\rangle_{2} \rightarrow\exp \left\{i \phi_{i j}\right\}|i\rangle_{1}|j\rangle_{2}$ \cite{leibfried2003experimental, PhysRevLett.75.4710,PhysRevLett.75.346,PhysRevLett.90.197902}. QPG becomes universal when $\phi=\phi_{11}+\phi_{00}-\phi_{10}-\phi_{01} \neq 0$. Under this operation, the quantum state of the system is evolved into the following one,
\begin{eqnarray}
    \begin{aligned}
        |\varphi_{2}\rangle_{ABC}=&P^{m,s}_{AC}|\varphi_{1}\rangle_{ABC}\\
                                 =&\sum_{m,n=0}^{d-1} \sum_{s=0}^{d-1} \lambda_{sn}|x_{s}|e^{i\theta_{s}}|snk_{m}\rangle_{ABC},
    \end{aligned} 
\end{eqnarray}

where
\begin{eqnarray}
    \begin{aligned}
       P^{m,s}_{AC}=\sum_{m=0}^{d-1} \sum_{s=0}^{d-1} e^{\mp i\theta_{m,s}}|sk_{m}\rangle \langle sk_{m}|.
    \end{aligned} 
\end{eqnarray}

\textbf{Step (\uppercase\expandafter{\romannumeral3})} For the sake of achieving DRSP without increasing quantum resource consumption, it is feasible to transfer qudit A from Alice to Bob. Communication security is ensured by encoding the prepared information in a non-local hyperentangled quantum state. Bob subsequently has the capability to perform joint operations on qudits A and B. This solution is also adopted in the experiment for a novel teleportation protocol that is in principle unconditional and requires only a single photon as an ex-anteprepared resource \cite{PhysRevLett.126.130502}.
\begin{eqnarray}
    \begin{aligned}
        |\varphi_{3}\rangle_{ABC}=&C^{s,n}_{AB}|\varphi_{2}\rangle_{ABC}\\
                                 =&\sum_{m,n=0}^{d-1} \sum_{s=0}^{d-1} \lambda_{s(n\oplus k_{s})}|x_{s}|e^{i\theta_{s}}|s(n\oplus k_{s})k_{m}\rangle_{ABC}\\
                                 =&\sum_{m,l=0}^{d-1} \sum_{s=0}^{d-1} \lambda_{sl}|x_{s}|e^{i\theta_{s}}|slk_{m}\rangle_{ABC}.
    \end{aligned} 
\end{eqnarray}

\textbf{Step (\uppercase\expandafter{\romannumeral4})} To reach the target further, Bob applies another C-NOT gate on qudits B and A. Following this procedure, only the process of measuring the irrelevant particles remains at the end.
\begin{eqnarray}
    \begin{aligned}
        |\varphi_{4}\rangle_{ABC}=&C^{l,s}_{BA}|\varphi_{3}\rangle_{ABC}\\
                                 =&\sum_{m,l=0}^{d-1} \sum_{s=0}^{d-1} \lambda_{(s\oplus k_{l})l}|x_{q}|e^{i\theta_{q}}|(s\oplus k_{l})lk_{m}\rangle_{ABC}\\
                                 =&\sum_{m,l=0}^{d-1} \sum_{q=0}^{d-1} \lambda_{ql}|x_{q}|e^{i\theta_{q}}|qlk_{m}\rangle_{ABC}\\
                                 =&\sum_{q,m=0}^{d-1}\lambda_{qm}|qm\rangle_{AC} \sum_{n=0}^{d-1}|x_{n}|e^{i\theta_{n}}|n\rangle_{B}.\label{equation 8}
    \end{aligned} 
\end{eqnarray}

\textbf{Step (\uppercase\expandafter{\romannumeral5})} Single-particle measuring is always easier to operationalize than joint measurement on multiple particles practically. For simplicity, two communicators execute PM on auxiliary qudits respectively, which is differentiated from previous RSP schemes as well. Communicators do not have to make joint Bell measurements implementable but challenging currently \cite{PhysRevA.59.3295, PhysRevA.99.023854,PhysRevA.99.052301, PhysRevLett.123.070505}. Instead, Alice and Bob implement PM on qudits C and A under the simplest basis \{0,1\} severally. RSP is always successful with either measurement result. Bob surely obtains the quantum state $x_{0}|0\rangle_{B}+x_{1}|1\rangle_{B}$ (Eq. (\ref{equation 1})) which Alice prepares for him.

The above represent the framework of the optimal DRSP protocol exhibited in this investigation. Appendix \ref{app2} displays the case of preparing a two-dimensional quantum state deterministically through a maximally entangled channel. One of the most significant highlights is the success probability is independent of the utilized entangled quantum channel, derived from Eq. (\ref{equation 8}). No matter how weak the entanglement of the quantum channel is, the success probability of RSP is always unit faithfully. In terms of reducing the experimental complexity, the requirements for preparing quantum channels are reduced remarkably, and a simple measurement approach is adopted in this DRSP protocol.
\begin{figure*}[htbp]
    \centering
    \includegraphics[width=0.8\paperwidth]{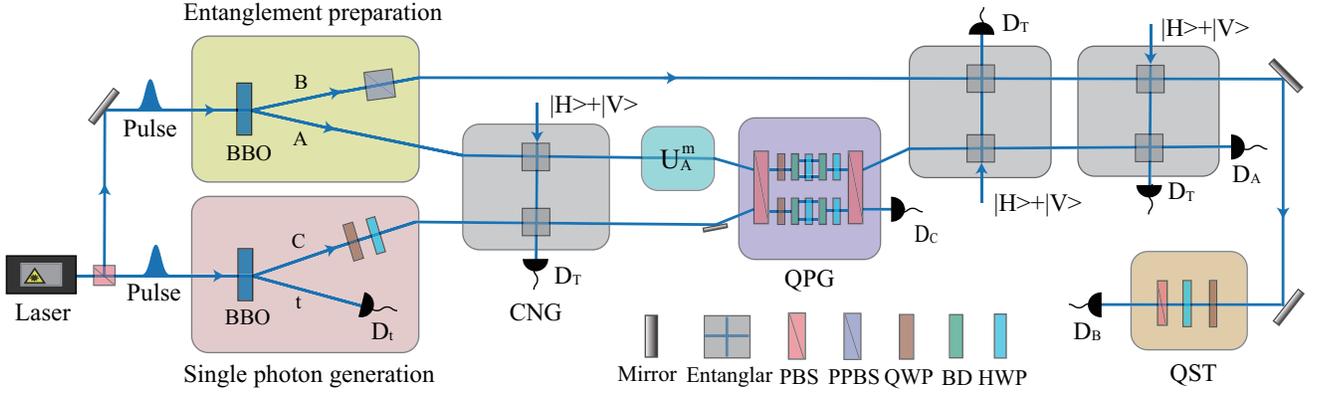}
    \caption{Experimental diagram to prepare a polarized state of a photon in a remote place via a non-maximally entangled channel. A pulsed ultraviolet (UV) laser beam is split to focus on two barium borate nonlinear (BBO) crystals and produces two pairs of polarization-entangled A-B and C-t. PPBS is employed to prepare a pure non-maximally entangled channel $|\varphi\rangle_{AB}=\alpha|HH\rangle+\beta|VV\rangle)_{AB}$. The heralded single-photon polarization state is set on $|H\rangle_{C}$, triggered by its twisted photon t. Alice firstly performs a C-NOT gate on photons A and C. One extra ancilla photon $|H\rangle+|V\rangle$ is needed beyond the control and target photons to realize a deterministic C-NOT operation. Secondly, Alice applies single-photon unitary transformation $U^{m}_{A}$ on photon A. Thirdly, She employs a QPG on photons A and C and sends A to Bob. Bob subsequently performs C-NOT gates $C^{s,n}_{AB}$ and $C^{l,s}_{BA}$ on photons A and B in succession. The last step of the optimal DRSP protocol is measurement. PM is implemented on photons A and C severally. The quantum state of photon B is analyzed via QST to test the quality of this DRSP scheme. Note here that CNG is the C-NOT gate, PBS is the polarizing beam splitter, PPBS is the partial polarizing beam splitter, QWP is the quarter-wave plate, BD is the beam displacer and HWP is the half-wave plate.}\label{Fig.2}
\end{figure*}

\section{realization}
The experimental schematic diagram for remotely preparing a polarization-encoded photon via a generally entangled channel is illustrated in Fig. \ref{Fig.2}. Suppose the assignment of Alice is to facilitate Bob in preparing a polarization state of a photon: $|\varphi\rangle=x_{0}|H\rangle+x_{1}|V\rangle$ (hither H indicates the horizontal linear polarization, V indicates the vertical linear polarization). The function of ultrafast femtosecond laser pulses is to pump spontaneous parametric down-conversion (SPDC) processes \cite{Takeuchi:01, PhysRevA.68.013804, Niu:08} to prepare two pairs of polarization-entangled photons A-B and C-t \cite{PhysRevLett.117.210502, PhysRevLett.125.230501, Im}. The initial polarized state $|H\rangle_{C}$ of photon C is produced by using the set of half-wave plates and quarter-wave plates. The other photon entangled with photon C, i.e., t, is detected at detector $D_{t}$, heralding the presence of a single photon in mode C. To confirm the generalization of our protocol, a partial polarizing beam splitter is placed in the path of photon B to produce a non-maximally entangled channel $|\varphi\rangle_{AB}=\alpha|HH\rangle+\beta|VV\rangle_{AB}$. After the preparation of the non-maximally quantum channel and an auxiliary polarized quantum state is completed, Alice performs a C-NOT gate on the control photon A and the target photon C. Note that one auxiliary photon set in the polarized state $|H\rangle+|V\rangle$ is craved for the performance of a faithful CNOT operation \cite{PhysRevLett.93.250502, PhysRevLett.94.030501, yamamoto2003demonstration}. This ancilla is unconsumed in the gate operation and can be recycled for further use. Because Alice knows the prepared quantum information for Bob, she possesses the capability to apply unitary operation $U^{m}_{A}$ associated with the information on photon A. This operation can be decomposed into a combination of Hadamard gates \cite{PhysRevA.84.050301, PhysRevA.87.012307, Larsen} and phase gates for more elementary operation in the experiment. Alice subsequently employs a QPG on photons A and C, which is easy to achieve with the current technology \cite{PhysRevLett.83.5166, PhysRevLett.106.013602, PhysRevLett.124.160501}. Afterward, photon A is sent to Bob, then two C-NOT gates with various control and target photons are employed on photons A and B. PM is implemented on photons A and C severally. The quality of preparation, which can be estimated by performing quantum state tomography (QST)  \cite{PhysRevLett.74.4101, PhysRevLett.105.150401, PhysRevLett.126.100402} between the prepared polarized state $|\varphi\rangle=x_{0}|H\rangle+x_{1}|V\rangle$ and the obtained polarized state of photon B.

\begin{figure}[htbp]
    \includegraphics[width=0.4\paperwidth]{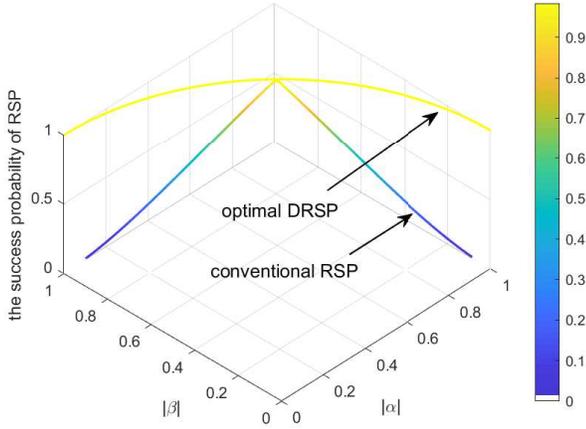}
    \caption{The success probability in the conventional RSP and optimal DRSP protocols is a function of entanglement coefficients of the quantum channel. As the quantum channel degenerates from a maximally entangled state $(|\alpha|=|\beta|=\frac{1}{\sqrt{2}})$ to a partially entangled one $(|\alpha|\neq |\beta|)$, the success probability of conventional RSP schemes is decayed accordingly, but the success probability of our optimal DRSP scheme remains 1 and constant.}\label{Fig.3}
\end{figure}

\section{discussion}
Herein we offer an alternative solution and demonstrate a novel DRSP protocol that allows for, in principle, increasing the success probability of RSP from $2|\alpha|^{2} (|\alpha|^{2}\leq\frac{1}{2})$ to 100\% without additional consumption of quantum resources. Remote preparation of quantum information encoded in quantum states demands entangled channels. The majority of attempts for improving the efficiency of state preparation are intended to enhance the entanglement strength of quantum channels by investing more quantum resources. The optimal DRSP protocol well-designed in this investigation is unconditional in improving entangled channels. As illustrated in Fig. \ref{Fig.3}, the probability of success is affected by the entanglement strength of the quantum channel in conventional RSP schemes. Nothing but when the quantum channel is a maximally entangled state, i.e., $|\alpha|=|\beta|$, the probability of success can reach 1. However, the success probability of the optimal DRSP scheme proposed in this study is independent of the entanglement strength of quantum channels. Even if the utilized quantum channel is degraded to a non-maximally entangled one, i.e., $|\alpha|\neq |\beta|$ the success probability of RSP remains unit and is always constant without increasing quantum resources consumption. Besides, this ingenious DRSP protocol only craves single communication,  whereas Roa's scheme \cite{PhysRevA.91.012344} needs multiple correspondences to transfer an equivalent amount of quantum information. It is unnecessary to construct multiple \cite{PhysRevA.73.022340} and repeated \cite{PhysRevA.91.012344} entanglement channels for the realization of deterministic communication assignments. In addition, we adopted a more convenient measurement method of PM instead of the complex approach of positive operator-valued measurement \cite{PhysRevA.98.042329}. Most previous RSP schemes are only suitable for the preparation of 2, 4, and 8-dimensional quantum states \cite{PhysRevA.65.022316}. Our generalized DRSP scheme applies to prepare quantum states of arbitrary dimensions. It is well-known that the ability of coherent manipulation of higher-dimensional quantum states is crucial for inventing advanced quantum technologies. Concerning the circumscribed two-dimensional systems, higher-dimensional systems possess virtues of noise resilience in quantum communications tasks and more efficient quantum computing and simulation. Consequently, the extension to remote preparation of high-dimensional quantum states realizable in this program is significant for developing quantum science and technology.

\section{conclusion}
We have addressed decoherence affecting remote preparation of quantum states and provided a noise-resistant faithful DRSP protocol. The sender prepares quantum information for the receiver deterministically in this neoteric scheme even if the quantum channel is a non-maximally entangled quantum state. The success possibility of RSP is boosted to 100\% without more consumption of quantum resources, which is beyond the standard protocol threshold. We are unconditional on heightening the entanglement of quantum channels as before. Moreover, our solution is more concise and realizable since we employ the generally entangled quantum channel and a simple measurement method concerning the previous RSP protocols. Our approach has provided a deeper insight into interesting extensions of traditional RSP schemes and boasts better performances in counteracting environmental noises in other quantum communication tasks. This study presents brilliant thinking for transferring information more efficiently in practical quantum networks.

\begin{center}
    \textbf{ACKNOWLEDGMENTS}
    \end{center}
We thank Yexiong Zeng and Chengsong Zhao for their fruitful discussions. This work was supported by National Natural Science Foundation of China (NSFC) under Grant No. 11574041.

\appendix

\section{Conventional preparation of a two-dimensional quantum state through a non-maximally entangled channel} \label{app1}
Suppose Alice aims to prepare a two-dimensional quantum state for Bob,
\begin{eqnarray}
    \begin{aligned}
        |\varphi\rangle&=x_{0}|0\rangle+x_{1}|1\rangle\\
                       &=x_{0}|0\rangle+|x_{1}|e^{i\theta}|1\rangle, \label{equation 11} 
    \end{aligned}
\end{eqnarray}
where $x_{0}$ is a real number, $x_{1}$ is a complex number, and both satisfy orthogonal normalization $x_{0}^{2}+|x_{1}|^{2}=1$. The generally entangled channel preshared by two communicators is the entangled quantum state of qubits A and B,
\begin{eqnarray}
    |\varphi\rangle_{AB}=(\alpha|00\rangle+\beta|11\rangle)_{AB}, \label{equation 12} 
\end{eqnarray}
where complex numbers $\alpha$ and $\beta$ satisfy orthogonal normalization $|\alpha|^{2}+|\beta|^{2}=1$ and $|\alpha|\leq|\beta|$. Only when $|\alpha|=|\beta|$, $|\varphi\rangle_{AB}$ is a maximally entangled quantum state. According to conventional RSP protocols \cite{nguyen2011remote, PhysRevA.98.042329}, Alice introduces an auxiliary qubit C with the
original state $|0\rangle_{C}$ and performs C-NOT gate $C_{AC}$ on qubits A and C,
\begin{eqnarray}
    \begin{aligned}
        |\varphi_0\rangle_{ABC}&=C_{AC}(|\varphi\rangle_{AB}\otimes|0\rangle_{C})\\
        &=(\alpha|000\rangle+\beta|111\rangle)_{ABC},
    \end{aligned} 
\end{eqnarray}
where
\begin{eqnarray}
    \begin{aligned}
       C_{ij}=&\prescript{}{i}(|0\rangle\langle 0|)_{i}\otimes\prescript{}{j}(|0\rangle\langle 0|+|1\rangle\langle 1|)_{j}\\
           &+\prescript{}{i}(|1\rangle\langle 1|)_{i}\otimes\prescript{}{j}(|1\rangle\langle 0|+|0\rangle\langle 1|)_{j}.      
    \end{aligned}
\end{eqnarray}
To carry out a general evolution, a collective unitary operation transforms the hyperentangled quantum state $(\alpha|000\rangle+\beta|111\rangle)_{ABC}$ to the result
\begin{eqnarray}
    \begin{aligned}
        |\varphi_1\rangle_{ABC}=&U_{AC}|\varphi_0\rangle_{ABC}\\
                               =&\alpha(|000\rangle+|111\rangle)_{ABC}\\
                               &+\sqrt{\beta^{2}-\alpha^{2}}|110\rangle_{ABC},      
    \end{aligned} 
\end{eqnarray}
where
\begin{eqnarray}
    \begin{aligned}
  U_{AC}=\left( \begin{array}{cccc}
    1 & 0 & 0 & 0 \\
    0 & 1 & 0 & 0 \\
    0 & 0 & \frac{\alpha}{\beta} & -\sqrt{1-\frac{\alpha^{2}}{\beta^{2}}} \\
    0 & 0 & \sqrt{1-\frac{\alpha^{2}}{\beta^{2}}} & \frac{\alpha}{\beta}
    \end{array} \right),
\end{aligned} 
\end{eqnarray}
here the agreement $|\alpha| \le |\beta|$. Alice then applies C-NOT gate $C_{AC}$ on qubits A and C again, the above quantum state is evoluted into
\begin{eqnarray}
    \begin{aligned}
        |\varphi_2\rangle_{ABC}=&C_{AC}|\varphi_1\rangle_{ABC}\\
                               =&\alpha|0\rangle_{C}(|00\rangle+|11\rangle)_{AB}\\
                               &+\sqrt{\beta^{2}-\alpha^{2}}|1\rangle_{C}|11\rangle_{AB}.     
    \end{aligned} 
\end{eqnarray}
Next, a measurement on the auxiliary qubit C follows. If the measurement result is $|0\rangle_{C}$, the three-qubit quantum state is collapsed into the maximally entangled quantum state $(|00\rangle+|11\rangle)_{AB}$ and RSP is successfully accessed. If the result
is $|1\rangle_{C}$, RSP fails with the disrupted quantum channel. Thereby the  successful probability of RSP may be expressed by $2|\alpha^{2}| (|\alpha^{2}| \le \frac{1}{2})$ according to traditional protocols.

\section{Deterministic preparation of a two-dimensional state via a non-maximally entangled channel} \label{app2}
The following section details the successful preparation of a two-dimensional quantum state (Eq. (\ref{equation 11})) at a distant location with one percent probability through a non-maximally entangled channel (Eq. (\ref{equation 12})).

\textbf{Step (\uppercase\expandafter{\romannumeral1})} Alice knows the information prepared for Bob applies the unitary transformation related to the information $U_{A}=(x_{0}|0\rangle+|x_{1}|e^{i\theta}|1\rangle)\langle 0|+(-|x_{1}|e^{-i\theta}|0\rangle+x_{0}|1\rangle)\langle 1|$ on qubit A. After that the hyperentangled state is unitarily evolved into
\begin{eqnarray}
 \begin{aligned} 
    |\varphi_1\rangle_{ABC}=&U_{A}|\varphi_0\rangle_{ABC}\\
                           =&(\alpha x_{0}|000\rangle+\alpha |x_{1}|e^{i\theta}|100\rangle\\
                             &-\beta |x_{1}|e^{-i\theta}|011\rangle+\beta x_{0}|111\rangle)_{ABC}.   
 \end{aligned}
\end{eqnarray}
Note that $U_{A}$ is an element of the two-dimensional special unitary group $SU_{2}$ constituted by a set of 2×2 complex matrices which have a determinant of unity. Any one of these elements can be generated by a set of generators \{$I,\sigma_{x},\sigma_{y},\sigma_{z}$\} \cite{PhysRevA.51.1015,PhysRevLett.74.4087,PhysRevA.52.3457}, i.e., $U_{A}$ can be substituted by these generators easy-to-prepare experimentally and corresponding operations.

\textbf{Step (\uppercase\expandafter{\romannumeral2})} Alice employs the phase gate $P_{AC}=\prescript{}{AC}(|00\rangle \langle 00|-e^{2i\theta}|01\rangle \langle 01|+|10\rangle \langle 10|+|11\rangle \langle 11|)_{AC}$ on qubits A and C. The quantum state of the system is evolved into the following one
\begin{eqnarray}
    \begin{aligned} 
        |\varphi_2\rangle_{ABC}=&P_{AC}|\varphi_1\rangle_{ABC}\\
        =&(\alpha x_{0}|000\rangle+\alpha |x_{1}|e^{i\theta}|100\rangle\\
        &+\beta |x_{1}|e^{i\theta}|011\rangle+\beta x_{0}|111\rangle)_{ABC}.      
\end{aligned} 
\end{eqnarray}

\textbf{Step (\uppercase\expandafter{\romannumeral3})} After particle transmission, Bob performs the C-NOT gate $C_{AB}$ on qubits A and B.
\begin{eqnarray}
    \begin{aligned} 
        |\varphi_3\rangle_{ABC}=&C_{AB}|\varphi_2\rangle_{ABC}\\
        =&(\alpha x_{0}|000\rangle+\alpha |x_{1}|e^{i\theta}|110\rangle\\
        &+\beta |x_{1}|e^{i\theta}|011\rangle+\beta x_{0}|101\rangle)_{ABC}.   
\end{aligned} 
\end{eqnarray}

\textbf{Step (\uppercase\expandafter{\romannumeral4})} Subsequently, Bob performs another C-NOT gate $C_{BA}$ on qubits B and A. 
\begin{eqnarray}
    \begin{aligned} 
        |\varphi_4\rangle_{ABC}=&C_{BA}|\varphi_3\rangle_{ABC}\\
        =&(\alpha x_{0}|000\rangle+\alpha |x_{1}|e^{i\theta}|010\rangle\\
        &+\beta |x_{1}|e^{-i\theta}|111\rangle+\beta x_{0}e^{-2i\theta}|101\rangle)_{ABC}\\
        =&\alpha|00\rangle_{AC}(x_{0}|0\rangle+|x_{1}|e^{i\theta}|1\rangle)_{B}\\
        &+\beta e^{-2i\theta}|11\rangle_{AC}(x_{0}|0\rangle+|x_{1}|e^{i\theta}|1\rangle)_{B}.
\end{aligned} 
\end{eqnarray}

\textbf{Step (\uppercase\expandafter{\romannumeral5})} Alice and Bob implement PM on qubits C and A under the simplest basis \{0,1\} severally. RSP is always successful with either measurement result. Bob surely obtains the quantum state (Eq. (\ref{equation 11})) prepared for him.

\bibliography{literature.bib} 

\begin{thebibliography}{86}%
\makeatletter
\providecommand \@ifxundefined [1]{%
 \@ifx{#1\undefined}
}%
\providecommand \@ifnum [1]{%
 \ifnum #1\expandafter \@firstoftwo
 \else \expandafter \@secondoftwo
 \fi
}%
\providecommand \@ifx [1]{%
 \ifx #1\expandafter \@firstoftwo
 \else \expandafter \@secondoftwo
 \fi
}%
\providecommand \natexlab [1]{#1}%
\providecommand \enquote  [1]{``#1''}%
\providecommand \bibnamefont  [1]{#1}%
\providecommand \bibfnamefont [1]{#1}%
\providecommand \citenamefont [1]{#1}%
\providecommand \href@noop [0]{\@secondoftwo}%
\providecommand \href [0]{\begingroup \@sanitize@url \@href}%
\providecommand \@href[1]{\@@startlink{#1}\@@href}%
\providecommand \@@href[1]{\endgroup#1\@@endlink}%
\providecommand \@sanitize@url [0]{\catcode `\\12\catcode `\$12\catcode
  `\&12\catcode `\#12\catcode `\^12\catcode `\_12\catcode `\%12\relax}%
\providecommand \@@startlink[1]{}%
\providecommand \@@endlink[0]{}%
\providecommand \url  [0]{\begingroup\@sanitize@url \@url }%
\providecommand \@url [1]{\endgroup\@href {#1}{\urlprefix }}%
\providecommand \urlprefix  [0]{URL }%
\providecommand \Eprint [0]{\href }%
\providecommand \doibase [0]{http://dx.doi.org/}%
\providecommand \selectlanguage [0]{\@gobble}%
\providecommand \bibinfo  [0]{\@secondoftwo}%
\providecommand \bibfield  [0]{\@secondoftwo}%
\providecommand \translation [1]{[#1]}%
\providecommand \BibitemOpen [0]{}%
\providecommand \bibitemStop [0]{}%
\providecommand \bibitemNoStop [0]{.\EOS\space}%
\providecommand \EOS [0]{\spacefactor3000\relax}%
\providecommand \BibitemShut  [1]{\csname bibitem#1\endcsname}%
\let\auto@bib@innerbib\@empty
\bibitem [{\citenamefont {Yurke}\ and\ \citenamefont
  {Denker}(1984)}]{PhysRevA.29.1419}%
  \BibitemOpen
  \bibfield  {author} {\bibinfo {author} {\bibfnamefont {B.}~\bibnamefont
  {Yurke}}\ and\ \bibinfo {author} {\bibfnamefont {J.~S.}\ \bibnamefont
  {Denker}},\ }\href {\doibase 10.1103/PhysRevA.29.1419} {\bibfield  {journal}
  {\bibinfo  {journal} {Phys. Rev. A}\ }\textbf {\bibinfo {volume} {29}},\
  \bibinfo {pages} {1419} (\bibinfo {year} {1984})}\BibitemShut {NoStop}%
\bibitem [{\citenamefont {Cirac}\ \emph {et~al.}(1997)\citenamefont {Cirac},
  \citenamefont {Zoller}, \citenamefont {Kimble},\ and\ \citenamefont
  {Mabuchi}}]{PhysRevLett.78.3221}%
  \BibitemOpen
  \bibfield  {author} {\bibinfo {author} {\bibfnamefont {J.~I.}\ \bibnamefont
  {Cirac}}, \bibinfo {author} {\bibfnamefont {P.}~\bibnamefont {Zoller}},
  \bibinfo {author} {\bibfnamefont {H.~J.}\ \bibnamefont {Kimble}}, \ and\
  \bibinfo {author} {\bibfnamefont {H.}~\bibnamefont {Mabuchi}},\ }\href
  {\doibase 10.1103/PhysRevLett.78.3221} {\bibfield  {journal} {\bibinfo
  {journal} {Phys. Rev. Lett.}\ }\textbf {\bibinfo {volume} {78}},\ \bibinfo
  {pages} {3221} (\bibinfo {year} {1997})}\BibitemShut {NoStop}%
\bibitem [{\citenamefont {Ritter}\ \emph {et~al.}(2012)\citenamefont {Ritter},
  \citenamefont {N{\"o}lleke}, \citenamefont {Hahn}, \citenamefont {Reiserer},
  \citenamefont {Neuzner}, \citenamefont {Uphoff}, \citenamefont {M{\"u}cke},
  \citenamefont {Figueroa}, \citenamefont {Bochmann},\ and\ \citenamefont
  {Rempe}}]{ritter2012elementary}%
  \BibitemOpen
  \bibfield  {author} {\bibinfo {author} {\bibfnamefont {S.}~\bibnamefont
  {Ritter}}, \bibinfo {author} {\bibfnamefont {C.}~\bibnamefont {N{\"o}lleke}},
  \bibinfo {author} {\bibfnamefont {C.}~\bibnamefont {Hahn}}, \bibinfo {author}
  {\bibfnamefont {A.}~\bibnamefont {Reiserer}}, \bibinfo {author}
  {\bibfnamefont {A.}~\bibnamefont {Neuzner}}, \bibinfo {author} {\bibfnamefont
  {M.}~\bibnamefont {Uphoff}}, \bibinfo {author} {\bibfnamefont
  {M.}~\bibnamefont {M{\"u}cke}}, \bibinfo {author} {\bibfnamefont
  {E.}~\bibnamefont {Figueroa}}, \bibinfo {author} {\bibfnamefont
  {J.}~\bibnamefont {Bochmann}}, \ and\ \bibinfo {author} {\bibfnamefont
  {G.}~\bibnamefont {Rempe}},\ }\href {\doibase 10.1038/nature11023} {\bibfield
   {journal} {\bibinfo  {journal} {Nature}\ }\textbf {\bibinfo {volume}
  {484}},\ \bibinfo {pages} {195} (\bibinfo {year} {2012})}\BibitemShut
  {NoStop}%
\bibitem [{\citenamefont {Simon}(2017)}]{simon2017towards}%
  \BibitemOpen
  \bibfield  {author} {\bibinfo {author} {\bibfnamefont {C.}~\bibnamefont
  {Simon}},\ }\href {\doibase 10.1038/s41566-017-0032-0} {\bibfield  {journal}
  {\bibinfo  {journal} {Nature Photonics}\ }\textbf {\bibinfo {volume} {11}},\
  \bibinfo {pages} {678} (\bibinfo {year} {2017})}\BibitemShut {NoStop}%
\bibitem [{\citenamefont {Liao}\ \emph {et~al.}(2018)\citenamefont {Liao},
  \citenamefont {Cai}, \citenamefont {Handsteiner}, \citenamefont {Liu},
  \citenamefont {Yin}, \citenamefont {Zhang}, \citenamefont {Rauch},
  \citenamefont {Fink}, \citenamefont {Ren}, \citenamefont {Liu}, \citenamefont
  {Li}, \citenamefont {Shen}, \citenamefont {Cao}, \citenamefont {Li},
  \citenamefont {Wang}, \citenamefont {Huang}, \citenamefont {Deng},
  \citenamefont {Xi}, \citenamefont {Ma}, \citenamefont {Hu}, \citenamefont
  {Li}, \citenamefont {Liu}, \citenamefont {Koidl}, \citenamefont {Wang},
  \citenamefont {Chen}, \citenamefont {Wang}, \citenamefont {Steindorfer},
  \citenamefont {Kirchner}, \citenamefont {Lu}, \citenamefont {Shu},
  \citenamefont {Ursin}, \citenamefont {Scheidl}, \citenamefont {Peng},
  \citenamefont {Wang}, \citenamefont {Zeilinger},\ and\ \citenamefont
  {Pan}}]{PhysRevLett.120.030501}%
  \BibitemOpen
  \bibfield  {author} {\bibinfo {author} {\bibfnamefont {S.-K.}\ \bibnamefont
  {Liao}}, \bibinfo {author} {\bibfnamefont {W.-Q.}\ \bibnamefont {Cai}},
  \bibinfo {author} {\bibfnamefont {J.}~\bibnamefont {Handsteiner}}, \bibinfo
  {author} {\bibfnamefont {B.}~\bibnamefont {Liu}}, \bibinfo {author}
  {\bibfnamefont {J.}~\bibnamefont {Yin}}, \bibinfo {author} {\bibfnamefont
  {L.}~\bibnamefont {Zhang}}, \bibinfo {author} {\bibfnamefont
  {D.}~\bibnamefont {Rauch}}, \bibinfo {author} {\bibfnamefont
  {M.}~\bibnamefont {Fink}}, \bibinfo {author} {\bibfnamefont {J.-G.}\
  \bibnamefont {Ren}}, \bibinfo {author} {\bibfnamefont {W.-Y.}\ \bibnamefont
  {Liu}}, \bibinfo {author} {\bibfnamefont {Y.}~\bibnamefont {Li}}, \bibinfo
  {author} {\bibfnamefont {Q.}~\bibnamefont {Shen}}, \bibinfo {author}
  {\bibfnamefont {Y.}~\bibnamefont {Cao}}, \bibinfo {author} {\bibfnamefont
  {F.-Z.}\ \bibnamefont {Li}}, \bibinfo {author} {\bibfnamefont {J.-F.}\
  \bibnamefont {Wang}}, \bibinfo {author} {\bibfnamefont {Y.-M.}\ \bibnamefont
  {Huang}}, \bibinfo {author} {\bibfnamefont {L.}~\bibnamefont {Deng}},
  \bibinfo {author} {\bibfnamefont {T.}~\bibnamefont {Xi}}, \bibinfo {author}
  {\bibfnamefont {L.}~\bibnamefont {Ma}}, \bibinfo {author} {\bibfnamefont
  {T.}~\bibnamefont {Hu}}, \bibinfo {author} {\bibfnamefont {L.}~\bibnamefont
  {Li}}, \bibinfo {author} {\bibfnamefont {N.-L.}\ \bibnamefont {Liu}},
  \bibinfo {author} {\bibfnamefont {F.}~\bibnamefont {Koidl}}, \bibinfo
  {author} {\bibfnamefont {P.}~\bibnamefont {Wang}}, \bibinfo {author}
  {\bibfnamefont {Y.-A.}\ \bibnamefont {Chen}}, \bibinfo {author}
  {\bibfnamefont {X.-B.}\ \bibnamefont {Wang}}, \bibinfo {author}
  {\bibfnamefont {M.}~\bibnamefont {Steindorfer}}, \bibinfo {author}
  {\bibfnamefont {G.}~\bibnamefont {Kirchner}}, \bibinfo {author}
  {\bibfnamefont {C.-Y.}\ \bibnamefont {Lu}}, \bibinfo {author} {\bibfnamefont
  {R.}~\bibnamefont {Shu}}, \bibinfo {author} {\bibfnamefont {R.}~\bibnamefont
  {Ursin}}, \bibinfo {author} {\bibfnamefont {T.}~\bibnamefont {Scheidl}},
  \bibinfo {author} {\bibfnamefont {C.-Z.}\ \bibnamefont {Peng}}, \bibinfo
  {author} {\bibfnamefont {J.-Y.}\ \bibnamefont {Wang}}, \bibinfo {author}
  {\bibfnamefont {A.}~\bibnamefont {Zeilinger}}, \ and\ \bibinfo {author}
  {\bibfnamefont {J.-W.}\ \bibnamefont {Pan}},\ }\href {\doibase
  10.1103/PhysRevLett.120.030501} {\bibfield  {journal} {\bibinfo  {journal}
  {Phys. Rev. Lett.}\ }\textbf {\bibinfo {volume} {120}},\ \bibinfo {pages}
  {030501} (\bibinfo {year} {2018})}\BibitemShut {NoStop}%
\bibitem [{\citenamefont {Lo}(2000)}]{PhysRevA.62.012313}%
  \BibitemOpen
  \bibfield  {author} {\bibinfo {author} {\bibfnamefont {H.-K.}\ \bibnamefont
  {Lo}},\ }\href {\doibase 10.1103/PhysRevA.62.012313} {\bibfield  {journal}
  {\bibinfo  {journal} {Phys. Rev. A}\ }\textbf {\bibinfo {volume} {62}},\
  \bibinfo {pages} {012313} (\bibinfo {year} {2000})}\BibitemShut {NoStop}%
\bibitem [{\citenamefont {Bennett}\ \emph {et~al.}(2001)\citenamefont
  {Bennett}, \citenamefont {DiVincenzo}, \citenamefont {Shor}, \citenamefont
  {Smolin}, \citenamefont {Terhal},\ and\ \citenamefont
  {Wootters}}]{PhysRevLett.87.077902}%
  \BibitemOpen
  \bibfield  {author} {\bibinfo {author} {\bibfnamefont {C.~H.}\ \bibnamefont
  {Bennett}}, \bibinfo {author} {\bibfnamefont {D.~P.}\ \bibnamefont
  {DiVincenzo}}, \bibinfo {author} {\bibfnamefont {P.~W.}\ \bibnamefont
  {Shor}}, \bibinfo {author} {\bibfnamefont {J.~A.}\ \bibnamefont {Smolin}},
  \bibinfo {author} {\bibfnamefont {B.~M.}\ \bibnamefont {Terhal}}, \ and\
  \bibinfo {author} {\bibfnamefont {W.~K.}\ \bibnamefont {Wootters}},\ }\href
  {\doibase 10.1103/PhysRevLett.87.077902} {\bibfield  {journal} {\bibinfo
  {journal} {Phys. Rev. Lett.}\ }\textbf {\bibinfo {volume} {87}},\ \bibinfo
  {pages} {077902} (\bibinfo {year} {2001})}\BibitemShut {NoStop}%
\bibitem [{\citenamefont {Pati}(2000{\natexlab{a}})}]{pati2000minimum}%
  \BibitemOpen
  \bibfield  {author} {\bibinfo {author} {\bibfnamefont {A.~K.}\ \bibnamefont
  {Pati}},\ }\href {\doibase 10.1103/PhysRevA.63.014302} {\bibfield  {journal}
  {\bibinfo  {journal} {Physical Review A}\ }\textbf {\bibinfo {volume} {63}},\
  \bibinfo {pages} {014302} (\bibinfo {year} {2000}{\natexlab{a}})}\BibitemShut
  {NoStop}%
\bibitem [{\citenamefont {Nguyen}\ \emph {et~al.}(2011)\citenamefont {Nguyen},
  \citenamefont {Cao}, \citenamefont {Nung},\ and\ \citenamefont
  {Kim}}]{nguyen2011remote}%
  \BibitemOpen
  \bibfield  {author} {\bibinfo {author} {\bibfnamefont {B.~A.}\ \bibnamefont
  {Nguyen}}, \bibinfo {author} {\bibfnamefont {T.~B.}\ \bibnamefont {Cao}},
  \bibinfo {author} {\bibfnamefont {V.~D.}\ \bibnamefont {Nung}}, \ and\
  \bibinfo {author} {\bibfnamefont {J.}~\bibnamefont {Kim}},\ }\href {\doibase
  10.1088/2043-6262/2/3/035009} {\bibfield  {journal} {\bibinfo  {journal}
  {Advances in Natural Sciences}\ }\textbf {\bibinfo {volume} {2}},\ \bibinfo
  {pages} {035009} (\bibinfo {year} {2011})}\BibitemShut {NoStop}%
\bibitem [{\citenamefont {Shor}(1995)}]{PhysRevA.52.R2493}%
  \BibitemOpen
  \bibfield  {author} {\bibinfo {author} {\bibfnamefont {P.~W.}\ \bibnamefont
  {Shor}},\ }\href {\doibase 10.1103/PhysRevA.52.R2493} {\bibfield  {journal}
  {\bibinfo  {journal} {Phys. Rev. A}\ }\textbf {\bibinfo {volume} {52}},\
  \bibinfo {pages} {R2493} (\bibinfo {year} {1995})}\BibitemShut {NoStop}%
\bibitem [{\citenamefont {Zurek}\ \emph {et~al.}(1993)\citenamefont {Zurek},
  \citenamefont {Habib},\ and\ \citenamefont {Paz}}]{PhysRevLett.70.1187}%
  \BibitemOpen
  \bibfield  {author} {\bibinfo {author} {\bibfnamefont {W.~H.}\ \bibnamefont
  {Zurek}}, \bibinfo {author} {\bibfnamefont {S.}~\bibnamefont {Habib}}, \ and\
  \bibinfo {author} {\bibfnamefont {J.~P.}\ \bibnamefont {Paz}},\ }\href
  {\doibase 10.1103/PhysRevLett.70.1187} {\bibfield  {journal} {\bibinfo
  {journal} {Phys. Rev. Lett.}\ }\textbf {\bibinfo {volume} {70}},\ \bibinfo
  {pages} {1187} (\bibinfo {year} {1993})}\BibitemShut {NoStop}%
\bibitem [{\citenamefont {Lidar}\ \emph {et~al.}(1998)\citenamefont {Lidar},
  \citenamefont {Chuang},\ and\ \citenamefont {Whaley}}]{PhysRevLett.81.2594}%
  \BibitemOpen
  \bibfield  {author} {\bibinfo {author} {\bibfnamefont {D.~A.}\ \bibnamefont
  {Lidar}}, \bibinfo {author} {\bibfnamefont {I.~L.}\ \bibnamefont {Chuang}}, \
  and\ \bibinfo {author} {\bibfnamefont {K.~B.}\ \bibnamefont {Whaley}},\
  }\href {\doibase 10.1103/PhysRevLett.81.2594} {\bibfield  {journal} {\bibinfo
   {journal} {Phys. Rev. Lett.}\ }\textbf {\bibinfo {volume} {81}},\ \bibinfo
  {pages} {2594} (\bibinfo {year} {1998})}\BibitemShut {NoStop}%
\bibitem [{\citenamefont {Schlosshauer}(2005)}]{RevModPhys.76.1267}%
  \BibitemOpen
  \bibfield  {author} {\bibinfo {author} {\bibfnamefont {M.}~\bibnamefont
  {Schlosshauer}},\ }\href {\doibase 10.1103/RevModPhys.76.1267} {\bibfield
  {journal} {\bibinfo  {journal} {Rev. Mod. Phys.}\ }\textbf {\bibinfo {volume}
  {76}},\ \bibinfo {pages} {1267} (\bibinfo {year} {2005})}\BibitemShut
  {NoStop}%
\bibitem [{\citenamefont {Xiang}\ \emph {et~al.}(2005)\citenamefont {Xiang},
  \citenamefont {Li}, \citenamefont {Yu},\ and\ \citenamefont
  {Guo}}]{PhysRevA.72.012315}%
  \BibitemOpen
  \bibfield  {author} {\bibinfo {author} {\bibfnamefont {G.-Y.}\ \bibnamefont
  {Xiang}}, \bibinfo {author} {\bibfnamefont {J.}~\bibnamefont {Li}}, \bibinfo
  {author} {\bibfnamefont {B.}~\bibnamefont {Yu}}, \ and\ \bibinfo {author}
  {\bibfnamefont {G.-C.}\ \bibnamefont {Guo}},\ }\href {\doibase
  10.1103/PhysRevA.72.012315} {\bibfield  {journal} {\bibinfo  {journal} {Phys.
  Rev. A}\ }\textbf {\bibinfo {volume} {72}},\ \bibinfo {pages} {012315}
  (\bibinfo {year} {2005})}\BibitemShut {NoStop}%
\bibitem [{\citenamefont {Kimble}(2008)}]{kimble2008quantum}%
  \BibitemOpen
  \bibfield  {author} {\bibinfo {author} {\bibfnamefont {H.~J.}\ \bibnamefont
  {Kimble}},\ }\href {\doibase 10.1038/nature07127} {\bibfield  {journal}
  {\bibinfo  {journal} {Nature}\ }\textbf {\bibinfo {volume} {453}},\ \bibinfo
  {pages} {1023} (\bibinfo {year} {2008})}\BibitemShut {NoStop}%
\bibitem [{\citenamefont {Pirandola}\ and\ \citenamefont
  {Braunstein}(2016)}]{pirandola2016physics}%
  \BibitemOpen
  \bibfield  {author} {\bibinfo {author} {\bibfnamefont {S.}~\bibnamefont
  {Pirandola}}\ and\ \bibinfo {author} {\bibfnamefont {S.~L.}\ \bibnamefont
  {Braunstein}},\ }\href {\doibase 10.1038/532169a} {\bibfield  {journal}
  {\bibinfo  {journal} {Nature News}\ }\textbf {\bibinfo {volume} {532}},\
  \bibinfo {pages} {169} (\bibinfo {year} {2016})}\BibitemShut {NoStop}%
\bibitem [{\citenamefont {Wehner}\ \emph {et~al.}(2018)\citenamefont {Wehner},
  \citenamefont {Elkouss},\ and\ \citenamefont {Hanson}}]{wehner2018quantum}%
  \BibitemOpen
  \bibfield  {author} {\bibinfo {author} {\bibfnamefont {S.}~\bibnamefont
  {Wehner}}, \bibinfo {author} {\bibfnamefont {D.}~\bibnamefont {Elkouss}}, \
  and\ \bibinfo {author} {\bibfnamefont {R.}~\bibnamefont {Hanson}},\ }\href
  {\doibase 10.1126/science.aam9288} {\bibfield  {journal} {\bibinfo  {journal}
  {Science}\ }\textbf {\bibinfo {volume} {362}} (\bibinfo {year} {2018}),\
  10.1126/science.aam9288}\BibitemShut {NoStop}%
\bibitem [{\citenamefont {Bennett}\ and\ \citenamefont
  {Wiesner}(1992)}]{PhysRevLett.69.2881}%
  \BibitemOpen
  \bibfield  {author} {\bibinfo {author} {\bibfnamefont {C.~H.}\ \bibnamefont
  {Bennett}}\ and\ \bibinfo {author} {\bibfnamefont {S.~J.}\ \bibnamefont
  {Wiesner}},\ }\href {\doibase 10.1103/PhysRevLett.69.2881} {\bibfield
  {journal} {\bibinfo  {journal} {Phys. Rev. Lett.}\ }\textbf {\bibinfo
  {volume} {69}},\ \bibinfo {pages} {2881} (\bibinfo {year}
  {1992})}\BibitemShut {NoStop}%
\bibitem [{\citenamefont {Bennett}\ \emph {et~al.}(1993)\citenamefont
  {Bennett}, \citenamefont {Brassard}, \citenamefont {Cr\'epeau}, \citenamefont
  {Jozsa}, \citenamefont {Peres},\ and\ \citenamefont
  {Wootters}}]{PhysRevLett.70.1895}%
  \BibitemOpen
  \bibfield  {author} {\bibinfo {author} {\bibfnamefont {C.~H.}\ \bibnamefont
  {Bennett}}, \bibinfo {author} {\bibfnamefont {G.}~\bibnamefont {Brassard}},
  \bibinfo {author} {\bibfnamefont {C.}~\bibnamefont {Cr\'epeau}}, \bibinfo
  {author} {\bibfnamefont {R.}~\bibnamefont {Jozsa}}, \bibinfo {author}
  {\bibfnamefont {A.}~\bibnamefont {Peres}}, \ and\ \bibinfo {author}
  {\bibfnamefont {W.~K.}\ \bibnamefont {Wootters}},\ }\href {\doibase
  10.1103/PhysRevLett.70.1895} {\bibfield  {journal} {\bibinfo  {journal}
  {Phys. Rev. Lett.}\ }\textbf {\bibinfo {volume} {70}},\ \bibinfo {pages}
  {1895} (\bibinfo {year} {1993})}\BibitemShut {NoStop}%
\bibitem [{\citenamefont {Shor}\ and\ \citenamefont
  {Preskill}(2000)}]{PhysRevLett.85.441}%
  \BibitemOpen
  \bibfield  {author} {\bibinfo {author} {\bibfnamefont {P.~W.}\ \bibnamefont
  {Shor}}\ and\ \bibinfo {author} {\bibfnamefont {J.}~\bibnamefont
  {Preskill}},\ }\href {\doibase 10.1103/PhysRevLett.85.441} {\bibfield
  {journal} {\bibinfo  {journal} {Phys. Rev. Lett.}\ }\textbf {\bibinfo
  {volume} {85}},\ \bibinfo {pages} {441} (\bibinfo {year} {2000})}\BibitemShut
  {NoStop}%
\bibitem [{\citenamefont {Scarani}\ \emph {et~al.}(2009)\citenamefont
  {Scarani}, \citenamefont {Bechmann-Pasquinucci}, \citenamefont {Cerf},
  \citenamefont {Du\ifmmode~\check{s}\else \v{s}\fi{}ek}, \citenamefont
  {L\"utkenhaus},\ and\ \citenamefont {Peev}}]{RevModPhys.81.1301}%
  \BibitemOpen
  \bibfield  {author} {\bibinfo {author} {\bibfnamefont {V.}~\bibnamefont
  {Scarani}}, \bibinfo {author} {\bibfnamefont {H.}~\bibnamefont
  {Bechmann-Pasquinucci}}, \bibinfo {author} {\bibfnamefont {N.~J.}\
  \bibnamefont {Cerf}}, \bibinfo {author} {\bibfnamefont {M.}~\bibnamefont
  {Du\ifmmode~\check{s}\else \v{s}\fi{}ek}}, \bibinfo {author} {\bibfnamefont
  {N.}~\bibnamefont {L\"utkenhaus}}, \ and\ \bibinfo {author} {\bibfnamefont
  {M.}~\bibnamefont {Peev}},\ }\href {\doibase 10.1103/RevModPhys.81.1301}
  {\bibfield  {journal} {\bibinfo  {journal} {Rev. Mod. Phys.}\ }\textbf
  {\bibinfo {volume} {81}},\ \bibinfo {pages} {1301} (\bibinfo {year}
  {2009})}\BibitemShut {NoStop}%
\bibitem [{\citenamefont {Pirandola}\ \emph {et~al.}(2015)\citenamefont
  {Pirandola}, \citenamefont {Eisert}, \citenamefont {Weedbrook}, \citenamefont
  {Furusawa},\ and\ \citenamefont {Braunstein}}]{pirandola2015advances}%
  \BibitemOpen
  \bibfield  {author} {\bibinfo {author} {\bibfnamefont {S.}~\bibnamefont
  {Pirandola}}, \bibinfo {author} {\bibfnamefont {J.}~\bibnamefont {Eisert}},
  \bibinfo {author} {\bibfnamefont {C.}~\bibnamefont {Weedbrook}}, \bibinfo
  {author} {\bibfnamefont {A.}~\bibnamefont {Furusawa}}, \ and\ \bibinfo
  {author} {\bibfnamefont {S.~L.}\ \bibnamefont {Braunstein}},\ }\href
  {\doibase 10.1038/nphoton.2015.154} {\bibfield  {journal} {\bibinfo
  {journal} {Nature photonics}\ }\textbf {\bibinfo {volume} {9}},\ \bibinfo
  {pages} {641} (\bibinfo {year} {2015})}\BibitemShut {NoStop}%
\bibitem [{\citenamefont {Daki{\'c}}\ \emph {et~al.}(2012)\citenamefont
  {Daki{\'c}}, \citenamefont {Lipp}, \citenamefont {Ma}, \citenamefont
  {Ringbauer}, \citenamefont {Kropatschek}, \citenamefont {Barz}, \citenamefont
  {Paterek}, \citenamefont {Vedral}, \citenamefont {Zeilinger}, \citenamefont
  {Brukner},\ and\ \citenamefont {Walther}}]{dakic2012quantum}%
  \BibitemOpen
  \bibfield  {author} {\bibinfo {author} {\bibfnamefont {B.}~\bibnamefont
  {Daki{\'c}}}, \bibinfo {author} {\bibfnamefont {Y.~O.}\ \bibnamefont {Lipp}},
  \bibinfo {author} {\bibfnamefont {X.}~\bibnamefont {Ma}}, \bibinfo {author}
  {\bibfnamefont {M.}~\bibnamefont {Ringbauer}}, \bibinfo {author}
  {\bibfnamefont {S.}~\bibnamefont {Kropatschek}}, \bibinfo {author}
  {\bibfnamefont {S.}~\bibnamefont {Barz}}, \bibinfo {author} {\bibfnamefont
  {T.}~\bibnamefont {Paterek}}, \bibinfo {author} {\bibfnamefont
  {V.}~\bibnamefont {Vedral}}, \bibinfo {author} {\bibfnamefont
  {A.}~\bibnamefont {Zeilinger}}, \bibinfo {author} {\bibfnamefont
  {{\v{C}}.}~\bibnamefont {Brukner}}, \ and\ \bibinfo {author} {\bibfnamefont
  {P.}~\bibnamefont {Walther}},\ }\href {\doibase 10.1038/nphys2377} {\bibfield
   {journal} {\bibinfo  {journal} {Nature Physics}\ }\textbf {\bibinfo {volume}
  {8}},\ \bibinfo {pages} {666} (\bibinfo {year} {2012})}\BibitemShut {NoStop}%
\bibitem [{\citenamefont {Monz}\ \emph {et~al.}(2011)\citenamefont {Monz},
  \citenamefont {Schindler}, \citenamefont {Barreiro}, \citenamefont {Chwalla},
  \citenamefont {Nigg}, \citenamefont {Coish}, \citenamefont {Harlander},
  \citenamefont {H\"ansel}, \citenamefont {Hennrich},\ and\ \citenamefont
  {Blatt}}]{PhysRevLett.106.130506}%
  \BibitemOpen
  \bibfield  {author} {\bibinfo {author} {\bibfnamefont {T.}~\bibnamefont
  {Monz}}, \bibinfo {author} {\bibfnamefont {P.}~\bibnamefont {Schindler}},
  \bibinfo {author} {\bibfnamefont {J.~T.}\ \bibnamefont {Barreiro}}, \bibinfo
  {author} {\bibfnamefont {M.}~\bibnamefont {Chwalla}}, \bibinfo {author}
  {\bibfnamefont {D.}~\bibnamefont {Nigg}}, \bibinfo {author} {\bibfnamefont
  {W.~A.}\ \bibnamefont {Coish}}, \bibinfo {author} {\bibfnamefont
  {M.}~\bibnamefont {Harlander}}, \bibinfo {author} {\bibfnamefont
  {W.}~\bibnamefont {H\"ansel}}, \bibinfo {author} {\bibfnamefont
  {M.}~\bibnamefont {Hennrich}}, \ and\ \bibinfo {author} {\bibfnamefont
  {R.}~\bibnamefont {Blatt}},\ }\href {\doibase 10.1103/PhysRevLett.106.130506}
  {\bibfield  {journal} {\bibinfo  {journal} {Phys. Rev. Lett.}\ }\textbf
  {\bibinfo {volume} {106}},\ \bibinfo {pages} {130506} (\bibinfo {year}
  {2011})}\BibitemShut {NoStop}%
\bibitem [{\citenamefont {Yao}\ \emph {et~al.}(2012)\citenamefont {Yao},
  \citenamefont {Wang}, \citenamefont {Xu}, \citenamefont {Lu}, \citenamefont
  {Pan}, \citenamefont {Bao}, \citenamefont {Peng}, \citenamefont {Lu},
  \citenamefont {Chen},\ and\ \citenamefont {Pan}}]{yao2012observation}%
  \BibitemOpen
  \bibfield  {author} {\bibinfo {author} {\bibfnamefont {X.-C.}\ \bibnamefont
  {Yao}}, \bibinfo {author} {\bibfnamefont {T.-X.}\ \bibnamefont {Wang}},
  \bibinfo {author} {\bibfnamefont {P.}~\bibnamefont {Xu}}, \bibinfo {author}
  {\bibfnamefont {H.}~\bibnamefont {Lu}}, \bibinfo {author} {\bibfnamefont
  {G.-S.}\ \bibnamefont {Pan}}, \bibinfo {author} {\bibfnamefont {X.-H.}\
  \bibnamefont {Bao}}, \bibinfo {author} {\bibfnamefont {C.-Z.}\ \bibnamefont
  {Peng}}, \bibinfo {author} {\bibfnamefont {C.-Y.}\ \bibnamefont {Lu}},
  \bibinfo {author} {\bibfnamefont {Y.-A.}\ \bibnamefont {Chen}}, \ and\
  \bibinfo {author} {\bibfnamefont {J.-W.}\ \bibnamefont {Pan}},\ }\href
  {\doibase 10.1038/nphoton.2011.354} {\bibfield  {journal} {\bibinfo
  {journal} {Nature photonics}\ }\textbf {\bibinfo {volume} {6}},\ \bibinfo
  {pages} {225} (\bibinfo {year} {2012})}\BibitemShut {NoStop}%
\bibitem [{\citenamefont {Huang}\ \emph {et~al.}(2011)\citenamefont {Huang},
  \citenamefont {Liu}, \citenamefont {Peng}, \citenamefont {Li}, \citenamefont
  {Li}, \citenamefont {Li},\ and\ \citenamefont {Guo}}]{huang2011experimental}%
  \BibitemOpen
  \bibfield  {author} {\bibinfo {author} {\bibfnamefont {Y.-F.}\ \bibnamefont
  {Huang}}, \bibinfo {author} {\bibfnamefont {B.-H.}\ \bibnamefont {Liu}},
  \bibinfo {author} {\bibfnamefont {L.}~\bibnamefont {Peng}}, \bibinfo {author}
  {\bibfnamefont {Y.-H.}\ \bibnamefont {Li}}, \bibinfo {author} {\bibfnamefont
  {L.}~\bibnamefont {Li}}, \bibinfo {author} {\bibfnamefont {C.-F.}\
  \bibnamefont {Li}}, \ and\ \bibinfo {author} {\bibfnamefont {G.-C.}\
  \bibnamefont {Guo}},\ }\href {\doibase 10.1038/ncomms1556} {\bibfield
  {journal} {\bibinfo  {journal} {Nature communications}\ }\textbf {\bibinfo
  {volume} {2}},\ \bibinfo {pages} {1} (\bibinfo {year} {2011})}\BibitemShut
  {NoStop}%
\bibitem [{\citenamefont {Wagenknecht}\ \emph {et~al.}(2010)\citenamefont
  {Wagenknecht}, \citenamefont {Li}, \citenamefont {Reingruber}, \citenamefont
  {Bao}, \citenamefont {Goebel}, \citenamefont {Chen}, \citenamefont {Zhang},
  \citenamefont {Chen},\ and\ \citenamefont
  {Pan}}]{wagenknecht2010experimental}%
  \BibitemOpen
  \bibfield  {author} {\bibinfo {author} {\bibfnamefont {C.}~\bibnamefont
  {Wagenknecht}}, \bibinfo {author} {\bibfnamefont {C.-M.}\ \bibnamefont {Li}},
  \bibinfo {author} {\bibfnamefont {A.}~\bibnamefont {Reingruber}}, \bibinfo
  {author} {\bibfnamefont {X.-H.}\ \bibnamefont {Bao}}, \bibinfo {author}
  {\bibfnamefont {A.}~\bibnamefont {Goebel}}, \bibinfo {author} {\bibfnamefont
  {Y.-A.}\ \bibnamefont {Chen}}, \bibinfo {author} {\bibfnamefont
  {Q.}~\bibnamefont {Zhang}}, \bibinfo {author} {\bibfnamefont
  {K.}~\bibnamefont {Chen}}, \ and\ \bibinfo {author} {\bibfnamefont {J.-W.}\
  \bibnamefont {Pan}},\ }\href {\doibase 10.1038/nphoton.2010.123} {\bibfield
  {journal} {\bibinfo  {journal} {Nature Photonics}\ }\textbf {\bibinfo
  {volume} {4}},\ \bibinfo {pages} {549} (\bibinfo {year} {2010})}\BibitemShut
  {NoStop}%
\bibitem [{\citenamefont {Zhang}\ \emph {et~al.}(2011)\citenamefont {Zhang},
  \citenamefont {Jin}, \citenamefont {Yang}, \citenamefont {Dai}, \citenamefont
  {Yang}, \citenamefont {Zhao}, \citenamefont {Rui}, \citenamefont {He},
  \citenamefont {Jiang}, \citenamefont {Yang} \emph
  {et~al.}}]{zhang2011preparation}%
  \BibitemOpen
  \bibfield  {author} {\bibinfo {author} {\bibfnamefont {H.}~\bibnamefont
  {Zhang}}, \bibinfo {author} {\bibfnamefont {X.-M.}\ \bibnamefont {Jin}},
  \bibinfo {author} {\bibfnamefont {J.}~\bibnamefont {Yang}}, \bibinfo {author}
  {\bibfnamefont {H.-N.}\ \bibnamefont {Dai}}, \bibinfo {author} {\bibfnamefont
  {S.-J.}\ \bibnamefont {Yang}}, \bibinfo {author} {\bibfnamefont {T.-M.}\
  \bibnamefont {Zhao}}, \bibinfo {author} {\bibfnamefont {J.}~\bibnamefont
  {Rui}}, \bibinfo {author} {\bibfnamefont {Y.}~\bibnamefont {He}}, \bibinfo
  {author} {\bibfnamefont {X.}~\bibnamefont {Jiang}}, \bibinfo {author}
  {\bibfnamefont {F.}~\bibnamefont {Yang}},  \emph {et~al.},\ }\href {\doibase
  10.1038/nphoton.2011.213} {\bibfield  {journal} {\bibinfo  {journal} {Nature
  Photonics}\ }\textbf {\bibinfo {volume} {5}},\ \bibinfo {pages} {628}
  (\bibinfo {year} {2011})}\BibitemShut {NoStop}%
\bibitem [{\citenamefont {Yin}\ \emph {et~al.}(2020)\citenamefont {Yin},
  \citenamefont {Li}, \citenamefont {Liao}, \citenamefont {Yang}, \citenamefont
  {Cao}, \citenamefont {Zhang}, \citenamefont {Ren}, \citenamefont {Cai},
  \citenamefont {Liu}, \citenamefont {Li} \emph
  {et~al.}}]{yin2020entanglement}%
  \BibitemOpen
  \bibfield  {author} {\bibinfo {author} {\bibfnamefont {J.}~\bibnamefont
  {Yin}}, \bibinfo {author} {\bibfnamefont {Y.-H.}\ \bibnamefont {Li}},
  \bibinfo {author} {\bibfnamefont {S.-K.}\ \bibnamefont {Liao}}, \bibinfo
  {author} {\bibfnamefont {M.}~\bibnamefont {Yang}}, \bibinfo {author}
  {\bibfnamefont {Y.}~\bibnamefont {Cao}}, \bibinfo {author} {\bibfnamefont
  {L.}~\bibnamefont {Zhang}}, \bibinfo {author} {\bibfnamefont {J.-G.}\
  \bibnamefont {Ren}}, \bibinfo {author} {\bibfnamefont {W.-Q.}\ \bibnamefont
  {Cai}}, \bibinfo {author} {\bibfnamefont {W.-Y.}\ \bibnamefont {Liu}},
  \bibinfo {author} {\bibfnamefont {S.-L.}\ \bibnamefont {Li}},  \emph
  {et~al.},\ }\href {\doibase 10.1038/s41586-020-2401-y} {\bibfield  {journal}
  {\bibinfo  {journal} {Nature}\ }\textbf {\bibinfo {volume} {582}},\ \bibinfo
  {pages} {501} (\bibinfo {year} {2020})}\BibitemShut {NoStop}%
\bibitem [{\citenamefont {Pati}(2000{\natexlab{b}})}]{PhysRevA.63.014302}%
  \BibitemOpen
  \bibfield  {author} {\bibinfo {author} {\bibfnamefont {A.~K.}\ \bibnamefont
  {Pati}},\ }\href {\doibase 10.1103/PhysRevA.63.014302} {\bibfield  {journal}
  {\bibinfo  {journal} {Phys. Rev. A}\ }\textbf {\bibinfo {volume} {63}},\
  \bibinfo {pages} {014302} (\bibinfo {year} {2000}{\natexlab{b}})}\BibitemShut
  {NoStop}%
\bibitem [{\citenamefont {Bennett}\ \emph
  {et~al.}(1996{\natexlab{a}})\citenamefont {Bennett}, \citenamefont
  {Brassard}, \citenamefont {Popescu}, \citenamefont {Schumacher},
  \citenamefont {Smolin},\ and\ \citenamefont {Wootters}}]{PhysRevLett.76.722}%
  \BibitemOpen
  \bibfield  {author} {\bibinfo {author} {\bibfnamefont {C.~H.}\ \bibnamefont
  {Bennett}}, \bibinfo {author} {\bibfnamefont {G.}~\bibnamefont {Brassard}},
  \bibinfo {author} {\bibfnamefont {S.}~\bibnamefont {Popescu}}, \bibinfo
  {author} {\bibfnamefont {B.}~\bibnamefont {Schumacher}}, \bibinfo {author}
  {\bibfnamefont {J.~A.}\ \bibnamefont {Smolin}}, \ and\ \bibinfo {author}
  {\bibfnamefont {W.~K.}\ \bibnamefont {Wootters}},\ }\href {\doibase
  10.1103/PhysRevLett.76.722} {\bibfield  {journal} {\bibinfo  {journal} {Phys.
  Rev. Lett.}\ }\textbf {\bibinfo {volume} {76}},\ \bibinfo {pages} {722}
  (\bibinfo {year} {1996}{\natexlab{a}})}\BibitemShut {NoStop}%
\bibitem [{\citenamefont {Deutsch}\ \emph {et~al.}(1996)\citenamefont
  {Deutsch}, \citenamefont {Ekert}, \citenamefont {Jozsa}, \citenamefont
  {Macchiavello}, \citenamefont {Popescu},\ and\ \citenamefont
  {Sanpera}}]{PhysRevLett.77.2818}%
  \BibitemOpen
  \bibfield  {author} {\bibinfo {author} {\bibfnamefont {D.}~\bibnamefont
  {Deutsch}}, \bibinfo {author} {\bibfnamefont {A.}~\bibnamefont {Ekert}},
  \bibinfo {author} {\bibfnamefont {R.}~\bibnamefont {Jozsa}}, \bibinfo
  {author} {\bibfnamefont {C.}~\bibnamefont {Macchiavello}}, \bibinfo {author}
  {\bibfnamefont {S.}~\bibnamefont {Popescu}}, \ and\ \bibinfo {author}
  {\bibfnamefont {A.}~\bibnamefont {Sanpera}},\ }\href {\doibase
  10.1103/PhysRevLett.77.2818} {\bibfield  {journal} {\bibinfo  {journal}
  {Phys. Rev. Lett.}\ }\textbf {\bibinfo {volume} {77}},\ \bibinfo {pages}
  {2818} (\bibinfo {year} {1996})}\BibitemShut {NoStop}%
\bibitem [{\citenamefont {Bennett}\ \emph
  {et~al.}(1996{\natexlab{b}})\citenamefont {Bennett}, \citenamefont
  {DiVincenzo}, \citenamefont {Smolin},\ and\ \citenamefont
  {Wootters}}]{PhysRevA.54.3824}%
  \BibitemOpen
  \bibfield  {author} {\bibinfo {author} {\bibfnamefont {C.~H.}\ \bibnamefont
  {Bennett}}, \bibinfo {author} {\bibfnamefont {D.~P.}\ \bibnamefont
  {DiVincenzo}}, \bibinfo {author} {\bibfnamefont {J.~A.}\ \bibnamefont
  {Smolin}}, \ and\ \bibinfo {author} {\bibfnamefont {W.~K.}\ \bibnamefont
  {Wootters}},\ }\href {\doibase 10.1103/PhysRevA.54.3824} {\bibfield
  {journal} {\bibinfo  {journal} {Phys. Rev. A}\ }\textbf {\bibinfo {volume}
  {54}},\ \bibinfo {pages} {3824} (\bibinfo {year}
  {1996}{\natexlab{b}})}\BibitemShut {NoStop}%
\bibitem [{\citenamefont {Bombin}\ and\ \citenamefont
  {Martin-Delgado}(2006)}]{PhysRevLett.97.180501}%
  \BibitemOpen
  \bibfield  {author} {\bibinfo {author} {\bibfnamefont {H.}~\bibnamefont
  {Bombin}}\ and\ \bibinfo {author} {\bibfnamefont {M.~A.}\ \bibnamefont
  {Martin-Delgado}},\ }\href {\doibase 10.1103/PhysRevLett.97.180501}
  {\bibfield  {journal} {\bibinfo  {journal} {Phys. Rev. Lett.}\ }\textbf
  {\bibinfo {volume} {97}},\ \bibinfo {pages} {180501} (\bibinfo {year}
  {2006})}\BibitemShut {NoStop}%
\bibitem [{\citenamefont {Zwerger}\ \emph {et~al.}(2013)\citenamefont
  {Zwerger}, \citenamefont {Briegel},\ and\ \citenamefont
  {D\"ur}}]{PhysRevLett.110.260503}%
  \BibitemOpen
  \bibfield  {author} {\bibinfo {author} {\bibfnamefont {M.}~\bibnamefont
  {Zwerger}}, \bibinfo {author} {\bibfnamefont {H.~J.}\ \bibnamefont
  {Briegel}}, \ and\ \bibinfo {author} {\bibfnamefont {W.}~\bibnamefont
  {D\"ur}},\ }\href {\doibase 10.1103/PhysRevLett.110.260503} {\bibfield
  {journal} {\bibinfo  {journal} {Phys. Rev. Lett.}\ }\textbf {\bibinfo
  {volume} {110}},\ \bibinfo {pages} {260503} (\bibinfo {year}
  {2013})}\BibitemShut {NoStop}%
\bibitem [{\citenamefont {Hu}\ \emph {et~al.}(2021)\citenamefont {Hu},
  \citenamefont {Huang}, \citenamefont {Sheng}, \citenamefont {Zhou},
  \citenamefont {Liu}, \citenamefont {Guo}, \citenamefont {Zhang},
  \citenamefont {Xing}, \citenamefont {Huang}, \citenamefont {Li},\ and\
  \citenamefont {Guo}}]{PhysRevLett.126.010503}%
  \BibitemOpen
  \bibfield  {author} {\bibinfo {author} {\bibfnamefont {X.-M.}\ \bibnamefont
  {Hu}}, \bibinfo {author} {\bibfnamefont {C.-X.}\ \bibnamefont {Huang}},
  \bibinfo {author} {\bibfnamefont {Y.-B.}\ \bibnamefont {Sheng}}, \bibinfo
  {author} {\bibfnamefont {L.}~\bibnamefont {Zhou}}, \bibinfo {author}
  {\bibfnamefont {B.-H.}\ \bibnamefont {Liu}}, \bibinfo {author} {\bibfnamefont
  {Y.}~\bibnamefont {Guo}}, \bibinfo {author} {\bibfnamefont {C.}~\bibnamefont
  {Zhang}}, \bibinfo {author} {\bibfnamefont {W.-B.}\ \bibnamefont {Xing}},
  \bibinfo {author} {\bibfnamefont {Y.-F.}\ \bibnamefont {Huang}}, \bibinfo
  {author} {\bibfnamefont {C.-F.}\ \bibnamefont {Li}}, \ and\ \bibinfo {author}
  {\bibfnamefont {G.-C.}\ \bibnamefont {Guo}},\ }\href {\doibase
  10.1103/PhysRevLett.126.010503} {\bibfield  {journal} {\bibinfo  {journal}
  {Phys. Rev. Lett.}\ }\textbf {\bibinfo {volume} {126}},\ \bibinfo {pages}
  {010503} (\bibinfo {year} {2021})}\BibitemShut {NoStop}%
\bibitem [{\citenamefont {Jonathan}\ and\ \citenamefont
  {Plenio}(1999)}]{jonathan1999entanglement}%
  \BibitemOpen
  \bibfield  {author} {\bibinfo {author} {\bibfnamefont {D.}~\bibnamefont
  {Jonathan}}\ and\ \bibinfo {author} {\bibfnamefont {M.~B.}\ \bibnamefont
  {Plenio}},\ }\href {\doibase 10.1103/PhysRevLett.83.3566} {\bibfield
  {journal} {\bibinfo  {journal} {Physical Review Letters}\ }\textbf {\bibinfo
  {volume} {83}},\ \bibinfo {pages} {3566} (\bibinfo {year}
  {1999})}\BibitemShut {NoStop}%
\bibitem [{\citenamefont {Daftuar}\ and\ \citenamefont
  {Klimesh}(2001)}]{PhysRevA.64.042314}%
  \BibitemOpen
  \bibfield  {author} {\bibinfo {author} {\bibfnamefont {S.}~\bibnamefont
  {Daftuar}}\ and\ \bibinfo {author} {\bibfnamefont {M.}~\bibnamefont
  {Klimesh}},\ }\href {\doibase 10.1103/PhysRevA.64.042314} {\bibfield
  {journal} {\bibinfo  {journal} {Phys. Rev. A}\ }\textbf {\bibinfo {volume}
  {64}},\ \bibinfo {pages} {042314} (\bibinfo {year} {2001})}\BibitemShut
  {NoStop}%
\bibitem [{\citenamefont {van Dam}\ and\ \citenamefont
  {Hayden}(2003)}]{PhysRevA.67.060302}%
  \BibitemOpen
  \bibfield  {author} {\bibinfo {author} {\bibfnamefont {W.}~\bibnamefont {van
  Dam}}\ and\ \bibinfo {author} {\bibfnamefont {P.}~\bibnamefont {Hayden}},\
  }\href {\doibase 10.1103/PhysRevA.67.060302} {\bibfield  {journal} {\bibinfo
  {journal} {Phys. Rev. A}\ }\textbf {\bibinfo {volume} {67}},\ \bibinfo
  {pages} {060302} (\bibinfo {year} {2003})}\BibitemShut {NoStop}%
\bibitem [{\citenamefont {Sanders}\ and\ \citenamefont
  {Gour}(2009)}]{PhysRevA.79.054302}%
  \BibitemOpen
  \bibfield  {author} {\bibinfo {author} {\bibfnamefont {Y.~R.}\ \bibnamefont
  {Sanders}}\ and\ \bibinfo {author} {\bibfnamefont {G.}~\bibnamefont {Gour}},\
  }\href {\doibase 10.1103/PhysRevA.79.054302} {\bibfield  {journal} {\bibinfo
  {journal} {Phys. Rev. A}\ }\textbf {\bibinfo {volume} {79}},\ \bibinfo
  {pages} {054302} (\bibinfo {year} {2009})}\BibitemShut {NoStop}%
\bibitem [{\citenamefont {Popescu}(1995)}]{PhysRevLett.74.2619}%
  \BibitemOpen
  \bibfield  {author} {\bibinfo {author} {\bibfnamefont {S.}~\bibnamefont
  {Popescu}},\ }\href {\doibase 10.1103/PhysRevLett.74.2619} {\bibfield
  {journal} {\bibinfo  {journal} {Phys. Rev. Lett.}\ }\textbf {\bibinfo
  {volume} {74}},\ \bibinfo {pages} {2619} (\bibinfo {year}
  {1995})}\BibitemShut {NoStop}%
\bibitem [{\citenamefont {Peres}(1996)}]{PhysRevA.54.2685}%
  \BibitemOpen
  \bibfield  {author} {\bibinfo {author} {\bibfnamefont {A.}~\bibnamefont
  {Peres}},\ }\href {\doibase 10.1103/PhysRevA.54.2685} {\bibfield  {journal}
  {\bibinfo  {journal} {Phys. Rev. A}\ }\textbf {\bibinfo {volume} {54}},\
  \bibinfo {pages} {2685} (\bibinfo {year} {1996})}\BibitemShut {NoStop}%
\bibitem [{\citenamefont {Masanes}(2006)}]{PhysRevLett.96.150501}%
  \BibitemOpen
  \bibfield  {author} {\bibinfo {author} {\bibfnamefont {L.}~\bibnamefont
  {Masanes}},\ }\href {\doibase 10.1103/PhysRevLett.96.150501} {\bibfield
  {journal} {\bibinfo  {journal} {Phys. Rev. Lett.}\ }\textbf {\bibinfo
  {volume} {96}},\ \bibinfo {pages} {150501} (\bibinfo {year}
  {2006})}\BibitemShut {NoStop}%
\bibitem [{\citenamefont {Masanes}\ \emph {et~al.}(2008)\citenamefont
  {Masanes}, \citenamefont {Liang},\ and\ \citenamefont
  {Doherty}}]{PhysRevLett.100.090403}%
  \BibitemOpen
  \bibfield  {author} {\bibinfo {author} {\bibfnamefont {L.}~\bibnamefont
  {Masanes}}, \bibinfo {author} {\bibfnamefont {Y.-C.}\ \bibnamefont {Liang}},
  \ and\ \bibinfo {author} {\bibfnamefont {A.~C.}\ \bibnamefont {Doherty}},\
  }\href {\doibase 10.1103/PhysRevLett.100.090403} {\bibfield  {journal}
  {\bibinfo  {journal} {Phys. Rev. Lett.}\ }\textbf {\bibinfo {volume} {100}},\
  \bibinfo {pages} {090403} (\bibinfo {year} {2008})}\BibitemShut {NoStop}%
\bibitem [{\citenamefont {Liang}\ \emph {et~al.}(2012)\citenamefont {Liang},
  \citenamefont {Masanes},\ and\ \citenamefont {Rosset}}]{PhysRevA.86.052115}%
  \BibitemOpen
  \bibfield  {author} {\bibinfo {author} {\bibfnamefont {Y.-C.}\ \bibnamefont
  {Liang}}, \bibinfo {author} {\bibfnamefont {L.}~\bibnamefont {Masanes}}, \
  and\ \bibinfo {author} {\bibfnamefont {D.}~\bibnamefont {Rosset}},\ }\href
  {\doibase 10.1103/PhysRevA.86.052115} {\bibfield  {journal} {\bibinfo
  {journal} {Phys. Rev. A}\ }\textbf {\bibinfo {volume} {86}},\ \bibinfo
  {pages} {052115} (\bibinfo {year} {2012})}\BibitemShut {NoStop}%
\bibitem [{\citenamefont {Riera-S\`abat}\ \emph {et~al.}(2021)\citenamefont
  {Riera-S\`abat}, \citenamefont {Sekatski}, \citenamefont {Pirker},\ and\
  \citenamefont {D\"ur}}]{PhysRevLett.127.040502}%
  \BibitemOpen
  \bibfield  {author} {\bibinfo {author} {\bibfnamefont {F.}~\bibnamefont
  {Riera-S\`abat}}, \bibinfo {author} {\bibfnamefont {P.}~\bibnamefont
  {Sekatski}}, \bibinfo {author} {\bibfnamefont {A.}~\bibnamefont {Pirker}}, \
  and\ \bibinfo {author} {\bibfnamefont {W.}~\bibnamefont {D\"ur}},\ }\href
  {\doibase 10.1103/PhysRevLett.127.040502} {\bibfield  {journal} {\bibinfo
  {journal} {Phys. Rev. Lett.}\ }\textbf {\bibinfo {volume} {127}},\ \bibinfo
  {pages} {040502} (\bibinfo {year} {2021})}\BibitemShut {NoStop}%
\bibitem [{\citenamefont {Lipka-Bartosik}\ and\ \citenamefont
  {Skrzypczyk}(2021)}]{PhysRevLett.127.080502}%
  \BibitemOpen
  \bibfield  {author} {\bibinfo {author} {\bibfnamefont {P.}~\bibnamefont
  {Lipka-Bartosik}}\ and\ \bibinfo {author} {\bibfnamefont {P.}~\bibnamefont
  {Skrzypczyk}},\ }\href {\doibase 10.1103/PhysRevLett.127.080502} {\bibfield
  {journal} {\bibinfo  {journal} {Phys. Rev. Lett.}\ }\textbf {\bibinfo
  {volume} {127}},\ \bibinfo {pages} {080502} (\bibinfo {year}
  {2021})}\BibitemShut {NoStop}%
\bibitem [{\citenamefont {Li}\ \emph {et~al.}(2021)\citenamefont {Li},
  \citenamefont {Fang}, \citenamefont {Zhang}, \citenamefont {Tabia},
  \citenamefont {Lu},\ and\ \citenamefont {Liang}}]{PhysRevResearch.3.023045}%
  \BibitemOpen
  \bibfield  {author} {\bibinfo {author} {\bibfnamefont {J.-Y.}\ \bibnamefont
  {Li}}, \bibinfo {author} {\bibfnamefont {X.-X.}\ \bibnamefont {Fang}},
  \bibinfo {author} {\bibfnamefont {T.}~\bibnamefont {Zhang}}, \bibinfo
  {author} {\bibfnamefont {G.~N.~M.}\ \bibnamefont {Tabia}}, \bibinfo {author}
  {\bibfnamefont {H.}~\bibnamefont {Lu}}, \ and\ \bibinfo {author}
  {\bibfnamefont {Y.-C.}\ \bibnamefont {Liang}},\ }\href {\doibase
  10.1103/PhysRevResearch.3.023045} {\bibfield  {journal} {\bibinfo  {journal}
  {Phys. Rev. Research}\ }\textbf {\bibinfo {volume} {3}},\ \bibinfo {pages}
  {023045} (\bibinfo {year} {2021})}\BibitemShut {NoStop}%
\bibitem [{\citenamefont {Ye}\ \emph {et~al.}(2004)\citenamefont {Ye},
  \citenamefont {Zhang},\ and\ \citenamefont {Guo}}]{PhysRevA.69.022310}%
  \BibitemOpen
  \bibfield  {author} {\bibinfo {author} {\bibfnamefont {M.-Y.}\ \bibnamefont
  {Ye}}, \bibinfo {author} {\bibfnamefont {Y.-S.}\ \bibnamefont {Zhang}}, \
  and\ \bibinfo {author} {\bibfnamefont {G.-C.}\ \bibnamefont {Guo}},\ }\href
  {\doibase 10.1103/PhysRevA.69.022310} {\bibfield  {journal} {\bibinfo
  {journal} {Phys. Rev. A}\ }\textbf {\bibinfo {volume} {69}},\ \bibinfo
  {pages} {022310} (\bibinfo {year} {2004})}\BibitemShut {NoStop}%
\bibitem [{\citenamefont {Zeng}\ and\ \citenamefont
  {Zhang}(2002)}]{PhysRevA.65.022316}%
  \BibitemOpen
  \bibfield  {author} {\bibinfo {author} {\bibfnamefont {B.}~\bibnamefont
  {Zeng}}\ and\ \bibinfo {author} {\bibfnamefont {P.}~\bibnamefont {Zhang}},\
  }\href {\doibase 10.1103/PhysRevA.65.022316} {\bibfield  {journal} {\bibinfo
  {journal} {Phys. Rev. A}\ }\textbf {\bibinfo {volume} {65}},\ \bibinfo
  {pages} {022316} (\bibinfo {year} {2002})}\BibitemShut {NoStop}%
\bibitem [{\citenamefont {An}\ \emph {et~al.}(2018)\citenamefont {An},
  \citenamefont {Dat},\ and\ \citenamefont {Kim}}]{PhysRevA.98.042329}%
  \BibitemOpen
  \bibfield  {author} {\bibinfo {author} {\bibfnamefont {N.~B.}\ \bibnamefont
  {An}}, \bibinfo {author} {\bibfnamefont {L.~T.}\ \bibnamefont {Dat}}, \ and\
  \bibinfo {author} {\bibfnamefont {J.}~\bibnamefont {Kim}},\ }\href {\doibase
  10.1103/PhysRevA.98.042329} {\bibfield  {journal} {\bibinfo  {journal} {Phys.
  Rev. A}\ }\textbf {\bibinfo {volume} {98}},\ \bibinfo {pages} {042329}
  (\bibinfo {year} {2018})}\BibitemShut {NoStop}%
\bibitem [{\citenamefont {Price}\ \emph {et~al.}(1999)\citenamefont {Price},
  \citenamefont {Somaroo}, \citenamefont {Dunlop}, \citenamefont {Havel},\ and\
  \citenamefont {Cory}}]{PhysRevA.60.2777}%
  \BibitemOpen
  \bibfield  {author} {\bibinfo {author} {\bibfnamefont {M.~D.}\ \bibnamefont
  {Price}}, \bibinfo {author} {\bibfnamefont {S.~S.}\ \bibnamefont {Somaroo}},
  \bibinfo {author} {\bibfnamefont {A.~E.}\ \bibnamefont {Dunlop}}, \bibinfo
  {author} {\bibfnamefont {T.~F.}\ \bibnamefont {Havel}}, \ and\ \bibinfo
  {author} {\bibfnamefont {D.~G.}\ \bibnamefont {Cory}},\ }\href {\doibase
  10.1103/PhysRevA.60.2777} {\bibfield  {journal} {\bibinfo  {journal} {Phys.
  Rev. A}\ }\textbf {\bibinfo {volume} {60}},\ \bibinfo {pages} {2777}
  (\bibinfo {year} {1999})}\BibitemShut {NoStop}%
\bibitem [{\citenamefont {Yoshikawa}\ \emph {et~al.}(2008)\citenamefont
  {Yoshikawa}, \citenamefont {Miwa}, \citenamefont {Huck}, \citenamefont
  {Andersen}, \citenamefont {van Loock},\ and\ \citenamefont
  {Furusawa}}]{PhysRevLett.101.250501}%
  \BibitemOpen
  \bibfield  {author} {\bibinfo {author} {\bibfnamefont {J.-i.}\ \bibnamefont
  {Yoshikawa}}, \bibinfo {author} {\bibfnamefont {Y.}~\bibnamefont {Miwa}},
  \bibinfo {author} {\bibfnamefont {A.}~\bibnamefont {Huck}}, \bibinfo {author}
  {\bibfnamefont {U.~L.}\ \bibnamefont {Andersen}}, \bibinfo {author}
  {\bibfnamefont {P.}~\bibnamefont {van Loock}}, \ and\ \bibinfo {author}
  {\bibfnamefont {A.}~\bibnamefont {Furusawa}},\ }\href {\doibase
  10.1103/PhysRevLett.101.250501} {\bibfield  {journal} {\bibinfo  {journal}
  {Phys. Rev. Lett.}\ }\textbf {\bibinfo {volume} {101}},\ \bibinfo {pages}
  {250501} (\bibinfo {year} {2008})}\BibitemShut {NoStop}%
\bibitem [{\citenamefont {Vallone}\ \emph {et~al.}(2008)\citenamefont
  {Vallone}, \citenamefont {Pomarico}, \citenamefont {De~Martini},\ and\
  \citenamefont {Mataloni}}]{PhysRevLett.100.160502}%
  \BibitemOpen
  \bibfield  {author} {\bibinfo {author} {\bibfnamefont {G.}~\bibnamefont
  {Vallone}}, \bibinfo {author} {\bibfnamefont {E.}~\bibnamefont {Pomarico}},
  \bibinfo {author} {\bibfnamefont {F.}~\bibnamefont {De~Martini}}, \ and\
  \bibinfo {author} {\bibfnamefont {P.}~\bibnamefont {Mataloni}},\ }\href
  {\doibase 10.1103/PhysRevLett.100.160502} {\bibfield  {journal} {\bibinfo
  {journal} {Phys. Rev. Lett.}\ }\textbf {\bibinfo {volume} {100}},\ \bibinfo
  {pages} {160502} (\bibinfo {year} {2008})}\BibitemShut {NoStop}%
\bibitem [{\citenamefont {Leibfried}\ \emph {et~al.}(2003)\citenamefont
  {Leibfried}, \citenamefont {DeMarco}, \citenamefont {Meyer}, \citenamefont
  {Lucas}, \citenamefont {Barrett}, \citenamefont {Britton}, \citenamefont
  {Itano}, \citenamefont {Jelenkovi{\'c}}, \citenamefont {Langer},
  \citenamefont {Rosenband} \emph {et~al.}}]{leibfried2003experimental}%
  \BibitemOpen
  \bibfield  {author} {\bibinfo {author} {\bibfnamefont {D.}~\bibnamefont
  {Leibfried}}, \bibinfo {author} {\bibfnamefont {B.}~\bibnamefont {DeMarco}},
  \bibinfo {author} {\bibfnamefont {V.}~\bibnamefont {Meyer}}, \bibinfo
  {author} {\bibfnamefont {D.}~\bibnamefont {Lucas}}, \bibinfo {author}
  {\bibfnamefont {M.}~\bibnamefont {Barrett}}, \bibinfo {author} {\bibfnamefont
  {J.}~\bibnamefont {Britton}}, \bibinfo {author} {\bibfnamefont {W.~M.}\
  \bibnamefont {Itano}}, \bibinfo {author} {\bibfnamefont {B.}~\bibnamefont
  {Jelenkovi{\'c}}}, \bibinfo {author} {\bibfnamefont {C.}~\bibnamefont
  {Langer}}, \bibinfo {author} {\bibfnamefont {T.}~\bibnamefont {Rosenband}},
  \emph {et~al.},\ }\href {\doibase 10.1038/nature01492} {\bibfield  {journal}
  {\bibinfo  {journal} {Nature}\ }\textbf {\bibinfo {volume} {422}},\ \bibinfo
  {pages} {412} (\bibinfo {year} {2003})}\BibitemShut {NoStop}%
\bibitem [{\citenamefont {Turchette}\ \emph {et~al.}(1995)\citenamefont
  {Turchette}, \citenamefont {Hood}, \citenamefont {Lange}, \citenamefont
  {Mabuchi},\ and\ \citenamefont {Kimble}}]{PhysRevLett.75.4710}%
  \BibitemOpen
  \bibfield  {author} {\bibinfo {author} {\bibfnamefont {Q.~A.}\ \bibnamefont
  {Turchette}}, \bibinfo {author} {\bibfnamefont {C.~J.}\ \bibnamefont {Hood}},
  \bibinfo {author} {\bibfnamefont {W.}~\bibnamefont {Lange}}, \bibinfo
  {author} {\bibfnamefont {H.}~\bibnamefont {Mabuchi}}, \ and\ \bibinfo
  {author} {\bibfnamefont {H.~J.}\ \bibnamefont {Kimble}},\ }\href {\doibase
  10.1103/PhysRevLett.75.4710} {\bibfield  {journal} {\bibinfo  {journal}
  {Phys. Rev. Lett.}\ }\textbf {\bibinfo {volume} {75}},\ \bibinfo {pages}
  {4710} (\bibinfo {year} {1995})}\BibitemShut {NoStop}%
\bibitem [{\citenamefont {Lloyd}(1995)}]{PhysRevLett.75.346}%
  \BibitemOpen
  \bibfield  {author} {\bibinfo {author} {\bibfnamefont {S.}~\bibnamefont
  {Lloyd}},\ }\href {\doibase 10.1103/PhysRevLett.75.346} {\bibfield  {journal}
  {\bibinfo  {journal} {Phys. Rev. Lett.}\ }\textbf {\bibinfo {volume} {75}},\
  \bibinfo {pages} {346} (\bibinfo {year} {1995})}\BibitemShut {NoStop}%
\bibitem [{\citenamefont {Ottaviani}\ \emph {et~al.}(2003)\citenamefont
  {Ottaviani}, \citenamefont {Vitali}, \citenamefont {Artoni}, \citenamefont
  {Cataliotti},\ and\ \citenamefont {Tombesi}}]{PhysRevLett.90.197902}%
  \BibitemOpen
  \bibfield  {author} {\bibinfo {author} {\bibfnamefont {C.}~\bibnamefont
  {Ottaviani}}, \bibinfo {author} {\bibfnamefont {D.}~\bibnamefont {Vitali}},
  \bibinfo {author} {\bibfnamefont {M.}~\bibnamefont {Artoni}}, \bibinfo
  {author} {\bibfnamefont {F.}~\bibnamefont {Cataliotti}}, \ and\ \bibinfo
  {author} {\bibfnamefont {P.}~\bibnamefont {Tombesi}},\ }\href {\doibase
  10.1103/PhysRevLett.90.197902} {\bibfield  {journal} {\bibinfo  {journal}
  {Phys. Rev. Lett.}\ }\textbf {\bibinfo {volume} {90}},\ \bibinfo {pages}
  {197902} (\bibinfo {year} {2003})}\BibitemShut {NoStop}%
\bibitem [{\citenamefont {Langenfeld}\ \emph {et~al.}(2021)\citenamefont
  {Langenfeld}, \citenamefont {Welte}, \citenamefont {Hartung}, \citenamefont
  {Daiss}, \citenamefont {Thomas}, \citenamefont {Morin}, \citenamefont
  {Distante},\ and\ \citenamefont {Rempe}}]{PhysRevLett.126.130502}%
  \BibitemOpen
  \bibfield  {author} {\bibinfo {author} {\bibfnamefont {S.}~\bibnamefont
  {Langenfeld}}, \bibinfo {author} {\bibfnamefont {S.}~\bibnamefont {Welte}},
  \bibinfo {author} {\bibfnamefont {L.}~\bibnamefont {Hartung}}, \bibinfo
  {author} {\bibfnamefont {S.}~\bibnamefont {Daiss}}, \bibinfo {author}
  {\bibfnamefont {P.}~\bibnamefont {Thomas}}, \bibinfo {author} {\bibfnamefont
  {O.}~\bibnamefont {Morin}}, \bibinfo {author} {\bibfnamefont
  {E.}~\bibnamefont {Distante}}, \ and\ \bibinfo {author} {\bibfnamefont
  {G.}~\bibnamefont {Rempe}},\ }\href {\doibase 10.1103/PhysRevLett.126.130502}
  {\bibfield  {journal} {\bibinfo  {journal} {Phys. Rev. Lett.}\ }\textbf
  {\bibinfo {volume} {126}},\ \bibinfo {pages} {130502} (\bibinfo {year}
  {2021})}\BibitemShut {NoStop}%
\bibitem [{\citenamefont {L\"utkenhaus}\ \emph {et~al.}(1999)\citenamefont
  {L\"utkenhaus}, \citenamefont {Calsamiglia},\ and\ \citenamefont
  {Suominen}}]{PhysRevA.59.3295}%
  \BibitemOpen
  \bibfield  {author} {\bibinfo {author} {\bibfnamefont {N.}~\bibnamefont
  {L\"utkenhaus}}, \bibinfo {author} {\bibfnamefont {J.}~\bibnamefont
  {Calsamiglia}}, \ and\ \bibinfo {author} {\bibfnamefont {K.-A.}\ \bibnamefont
  {Suominen}},\ }\href {\doibase 10.1103/PhysRevA.59.3295} {\bibfield
  {journal} {\bibinfo  {journal} {Phys. Rev. A}\ }\textbf {\bibinfo {volume}
  {59}},\ \bibinfo {pages} {3295} (\bibinfo {year} {1999})}\BibitemShut
  {NoStop}%
\bibitem [{\citenamefont {Gonz\'alez-Guti\'errez}\ and\ \citenamefont
  {Torres}(2019)}]{PhysRevA.99.023854}%
  \BibitemOpen
  \bibfield  {author} {\bibinfo {author} {\bibfnamefont {C.~A.}\ \bibnamefont
  {Gonz\'alez-Guti\'errez}}\ and\ \bibinfo {author} {\bibfnamefont {J.~M.}\
  \bibnamefont {Torres}},\ }\href {\doibase 10.1103/PhysRevA.99.023854}
  {\bibfield  {journal} {\bibinfo  {journal} {Phys. Rev. A}\ }\textbf {\bibinfo
  {volume} {99}},\ \bibinfo {pages} {023854} (\bibinfo {year}
  {2019})}\BibitemShut {NoStop}%
\bibitem [{\citenamefont {Zhang}\ \emph {et~al.}(2019)\citenamefont {Zhang},
  \citenamefont {Zhang}, \citenamefont {Hu}, \citenamefont {Liu}, \citenamefont
  {Huang}, \citenamefont {Li},\ and\ \citenamefont {Guo}}]{PhysRevA.99.052301}%
  \BibitemOpen
  \bibfield  {author} {\bibinfo {author} {\bibfnamefont {H.}~\bibnamefont
  {Zhang}}, \bibinfo {author} {\bibfnamefont {C.}~\bibnamefont {Zhang}},
  \bibinfo {author} {\bibfnamefont {X.-M.}\ \bibnamefont {Hu}}, \bibinfo
  {author} {\bibfnamefont {B.-H.}\ \bibnamefont {Liu}}, \bibinfo {author}
  {\bibfnamefont {Y.-F.}\ \bibnamefont {Huang}}, \bibinfo {author}
  {\bibfnamefont {C.-F.}\ \bibnamefont {Li}}, \ and\ \bibinfo {author}
  {\bibfnamefont {G.-C.}\ \bibnamefont {Guo}},\ }\href {\doibase
  10.1103/PhysRevA.99.052301} {\bibfield  {journal} {\bibinfo  {journal} {Phys.
  Rev. A}\ }\textbf {\bibinfo {volume} {99}},\ \bibinfo {pages} {052301}
  (\bibinfo {year} {2019})}\BibitemShut {NoStop}%
\bibitem [{\citenamefont {Luo}\ \emph {et~al.}(2019)\citenamefont {Luo},
  \citenamefont {Zhong}, \citenamefont {Erhard}, \citenamefont {Wang},
  \citenamefont {Peng}, \citenamefont {Krenn}, \citenamefont {Jiang},
  \citenamefont {Li}, \citenamefont {Liu}, \citenamefont {Lu}, \citenamefont
  {Zeilinger},\ and\ \citenamefont {Pan}}]{PhysRevLett.123.070505}%
  \BibitemOpen
  \bibfield  {author} {\bibinfo {author} {\bibfnamefont {Y.-H.}\ \bibnamefont
  {Luo}}, \bibinfo {author} {\bibfnamefont {H.-S.}\ \bibnamefont {Zhong}},
  \bibinfo {author} {\bibfnamefont {M.}~\bibnamefont {Erhard}}, \bibinfo
  {author} {\bibfnamefont {X.-L.}\ \bibnamefont {Wang}}, \bibinfo {author}
  {\bibfnamefont {L.-C.}\ \bibnamefont {Peng}}, \bibinfo {author}
  {\bibfnamefont {M.}~\bibnamefont {Krenn}}, \bibinfo {author} {\bibfnamefont
  {X.}~\bibnamefont {Jiang}}, \bibinfo {author} {\bibfnamefont
  {L.}~\bibnamefont {Li}}, \bibinfo {author} {\bibfnamefont {N.-L.}\
  \bibnamefont {Liu}}, \bibinfo {author} {\bibfnamefont {C.-Y.}\ \bibnamefont
  {Lu}}, \bibinfo {author} {\bibfnamefont {A.}~\bibnamefont {Zeilinger}}, \
  and\ \bibinfo {author} {\bibfnamefont {J.-W.}\ \bibnamefont {Pan}},\ }\href
  {\doibase 10.1103/PhysRevLett.123.070505} {\bibfield  {journal} {\bibinfo
  {journal} {Phys. Rev. Lett.}\ }\textbf {\bibinfo {volume} {123}},\ \bibinfo
  {pages} {070505} (\bibinfo {year} {2019})}\BibitemShut {NoStop}%
\bibitem [{\citenamefont {Takeuchi}(2001)}]{Takeuchi:01}%
  \BibitemOpen
  \bibfield  {author} {\bibinfo {author} {\bibfnamefont {S.}~\bibnamefont
  {Takeuchi}},\ }\href {\doibase 10.1364/OL.26.000843} {\bibfield  {journal}
  {\bibinfo  {journal} {Opt. Lett.}\ }\textbf {\bibinfo {volume} {26}},\
  \bibinfo {pages} {843} (\bibinfo {year} {2001})}\BibitemShut {NoStop}%
\bibitem [{\citenamefont {Kim}(2003)}]{PhysRevA.68.013804}%
  \BibitemOpen
  \bibfield  {author} {\bibinfo {author} {\bibfnamefont {Y.-H.}\ \bibnamefont
  {Kim}},\ }\href {\doibase 10.1103/PhysRevA.68.013804} {\bibfield  {journal}
  {\bibinfo  {journal} {Phys. Rev. A}\ }\textbf {\bibinfo {volume} {68}},\
  \bibinfo {pages} {013804} (\bibinfo {year} {2003})}\BibitemShut {NoStop}%
\bibitem [{\citenamefont {Niu}\ \emph {et~al.}(2008)\citenamefont {Niu},
  \citenamefont {Huang}, \citenamefont {Xiang}, \citenamefont {Guo},\ and\
  \citenamefont {Ou}}]{Niu:08}%
  \BibitemOpen
  \bibfield  {author} {\bibinfo {author} {\bibfnamefont {X.-L.}\ \bibnamefont
  {Niu}}, \bibinfo {author} {\bibfnamefont {Y.-F.}\ \bibnamefont {Huang}},
  \bibinfo {author} {\bibfnamefont {G.-Y.}\ \bibnamefont {Xiang}}, \bibinfo
  {author} {\bibfnamefont {G.-C.}\ \bibnamefont {Guo}}, \ and\ \bibinfo
  {author} {\bibfnamefont {Z.~Y.}\ \bibnamefont {Ou}},\ }\href {\doibase
  10.1364/OL.33.000968} {\bibfield  {journal} {\bibinfo  {journal} {Opt.
  Lett.}\ }\textbf {\bibinfo {volume} {33}},\ \bibinfo {pages} {968} (\bibinfo
  {year} {2008})}\BibitemShut {NoStop}%
\bibitem [{\citenamefont {Wang}\ \emph {et~al.}(2016)\citenamefont {Wang},
  \citenamefont {Chen}, \citenamefont {Li}, \citenamefont {Huang},
  \citenamefont {Liu}, \citenamefont {Chen}, \citenamefont {Luo}, \citenamefont
  {Su}, \citenamefont {Wu}, \citenamefont {Li}, \citenamefont {Lu},
  \citenamefont {Hu}, \citenamefont {Jiang}, \citenamefont {Peng},
  \citenamefont {Li}, \citenamefont {Liu}, \citenamefont {Chen}, \citenamefont
  {Lu},\ and\ \citenamefont {Pan}}]{PhysRevLett.117.210502}%
  \BibitemOpen
  \bibfield  {author} {\bibinfo {author} {\bibfnamefont {X.-L.}\ \bibnamefont
  {Wang}}, \bibinfo {author} {\bibfnamefont {L.-K.}\ \bibnamefont {Chen}},
  \bibinfo {author} {\bibfnamefont {W.}~\bibnamefont {Li}}, \bibinfo {author}
  {\bibfnamefont {H.-L.}\ \bibnamefont {Huang}}, \bibinfo {author}
  {\bibfnamefont {C.}~\bibnamefont {Liu}}, \bibinfo {author} {\bibfnamefont
  {C.}~\bibnamefont {Chen}}, \bibinfo {author} {\bibfnamefont {Y.-H.}\
  \bibnamefont {Luo}}, \bibinfo {author} {\bibfnamefont {Z.-E.}\ \bibnamefont
  {Su}}, \bibinfo {author} {\bibfnamefont {D.}~\bibnamefont {Wu}}, \bibinfo
  {author} {\bibfnamefont {Z.-D.}\ \bibnamefont {Li}}, \bibinfo {author}
  {\bibfnamefont {H.}~\bibnamefont {Lu}}, \bibinfo {author} {\bibfnamefont
  {Y.}~\bibnamefont {Hu}}, \bibinfo {author} {\bibfnamefont {X.}~\bibnamefont
  {Jiang}}, \bibinfo {author} {\bibfnamefont {C.-Z.}\ \bibnamefont {Peng}},
  \bibinfo {author} {\bibfnamefont {L.}~\bibnamefont {Li}}, \bibinfo {author}
  {\bibfnamefont {N.-L.}\ \bibnamefont {Liu}}, \bibinfo {author} {\bibfnamefont
  {Y.-A.}\ \bibnamefont {Chen}}, \bibinfo {author} {\bibfnamefont {C.-Y.}\
  \bibnamefont {Lu}}, \ and\ \bibinfo {author} {\bibfnamefont {J.-W.}\
  \bibnamefont {Pan}},\ }\href {\doibase 10.1103/PhysRevLett.117.210502}
  {\bibfield  {journal} {\bibinfo  {journal} {Phys. Rev. Lett.}\ }\textbf
  {\bibinfo {volume} {117}},\ \bibinfo {pages} {210502} (\bibinfo {year}
  {2016})}\BibitemShut {NoStop}%
\bibitem [{\citenamefont {Hu}\ \emph {et~al.}(2020)\citenamefont {Hu},
  \citenamefont {Zhang}, \citenamefont {Liu}, \citenamefont {Cai},
  \citenamefont {Ye}, \citenamefont {Guo}, \citenamefont {Xing}, \citenamefont
  {Huang}, \citenamefont {Huang}, \citenamefont {Li},\ and\ \citenamefont
  {Guo}}]{PhysRevLett.125.230501}%
  \BibitemOpen
  \bibfield  {author} {\bibinfo {author} {\bibfnamefont {X.-M.}\ \bibnamefont
  {Hu}}, \bibinfo {author} {\bibfnamefont {C.}~\bibnamefont {Zhang}}, \bibinfo
  {author} {\bibfnamefont {B.-H.}\ \bibnamefont {Liu}}, \bibinfo {author}
  {\bibfnamefont {Y.}~\bibnamefont {Cai}}, \bibinfo {author} {\bibfnamefont
  {X.-J.}\ \bibnamefont {Ye}}, \bibinfo {author} {\bibfnamefont
  {Y.}~\bibnamefont {Guo}}, \bibinfo {author} {\bibfnamefont {W.-B.}\
  \bibnamefont {Xing}}, \bibinfo {author} {\bibfnamefont {C.-X.}\ \bibnamefont
  {Huang}}, \bibinfo {author} {\bibfnamefont {Y.-F.}\ \bibnamefont {Huang}},
  \bibinfo {author} {\bibfnamefont {C.-F.}\ \bibnamefont {Li}}, \ and\ \bibinfo
  {author} {\bibfnamefont {G.-C.}\ \bibnamefont {Guo}},\ }\href {\doibase
  10.1103/PhysRevLett.125.230501} {\bibfield  {journal} {\bibinfo  {journal}
  {Phys. Rev. Lett.}\ }\textbf {\bibinfo {volume} {125}},\ \bibinfo {pages}
  {230501} (\bibinfo {year} {2020})}\BibitemShut {NoStop}%
\bibitem [{\citenamefont {Im}\ \emph {et~al.}(2021)\citenamefont {Im},
  \citenamefont {Lee}, \citenamefont {Kim}, \citenamefont {Nha}, \citenamefont
  {Kim}, \citenamefont {Lee},\ and\ \citenamefont {Kim}}]{Im}%
  \BibitemOpen
  \bibfield  {author} {\bibinfo {author} {\bibfnamefont {D.-G.}\ \bibnamefont
  {Im}}, \bibinfo {author} {\bibfnamefont {C.-H.}\ \bibnamefont {Lee}},
  \bibinfo {author} {\bibfnamefont {Y.}~\bibnamefont {Kim}}, \bibinfo {author}
  {\bibfnamefont {H.}~\bibnamefont {Nha}}, \bibinfo {author} {\bibfnamefont
  {M.~S.}\ \bibnamefont {Kim}}, \bibinfo {author} {\bibfnamefont {S.-W.}\
  \bibnamefont {Lee}}, \ and\ \bibinfo {author} {\bibfnamefont {Y.-H.}\
  \bibnamefont {Kim}},\ }\href {\doibase 10.1038/s41534-021-00426-x} {\bibfield
   {journal} {\bibinfo  {journal} {npj Quantum Information}\ }\textbf {\bibinfo
  {volume} {7}},\ \bibinfo {pages} {86} (\bibinfo {year} {2021})}\BibitemShut
  {NoStop}%
\bibitem [{\citenamefont {Nemoto}\ and\ \citenamefont
  {Munro}(2004)}]{PhysRevLett.93.250502}%
  \BibitemOpen
  \bibfield  {author} {\bibinfo {author} {\bibfnamefont {K.}~\bibnamefont
  {Nemoto}}\ and\ \bibinfo {author} {\bibfnamefont {W.~J.}\ \bibnamefont
  {Munro}},\ }\href {\doibase 10.1103/PhysRevLett.93.250502} {\bibfield
  {journal} {\bibinfo  {journal} {Phys. Rev. Lett.}\ }\textbf {\bibinfo
  {volume} {93}},\ \bibinfo {pages} {250502} (\bibinfo {year}
  {2004})}\BibitemShut {NoStop}%
\bibitem [{\citenamefont {Zhao}\ \emph {et~al.}(2005)\citenamefont {Zhao},
  \citenamefont {Zhang}, \citenamefont {Chen}, \citenamefont {Zhang},
  \citenamefont {Du}, \citenamefont {Yang},\ and\ \citenamefont
  {Pan}}]{PhysRevLett.94.030501}%
  \BibitemOpen
  \bibfield  {author} {\bibinfo {author} {\bibfnamefont {Z.}~\bibnamefont
  {Zhao}}, \bibinfo {author} {\bibfnamefont {A.-N.}\ \bibnamefont {Zhang}},
  \bibinfo {author} {\bibfnamefont {Y.-A.}\ \bibnamefont {Chen}}, \bibinfo
  {author} {\bibfnamefont {H.}~\bibnamefont {Zhang}}, \bibinfo {author}
  {\bibfnamefont {J.-F.}\ \bibnamefont {Du}}, \bibinfo {author} {\bibfnamefont
  {T.}~\bibnamefont {Yang}}, \ and\ \bibinfo {author} {\bibfnamefont {J.-W.}\
  \bibnamefont {Pan}},\ }\href {\doibase 10.1103/PhysRevLett.94.030501}
  {\bibfield  {journal} {\bibinfo  {journal} {Phys. Rev. Lett.}\ }\textbf
  {\bibinfo {volume} {94}},\ \bibinfo {pages} {030501} (\bibinfo {year}
  {2005})}\BibitemShut {NoStop}%
\bibitem [{\citenamefont {Yamamoto}\ \emph {et~al.}(2003)\citenamefont
  {Yamamoto}, \citenamefont {Pashkin}, \citenamefont {Astafiev}, \citenamefont
  {Nakamura},\ and\ \citenamefont {Tsai}}]{yamamoto2003demonstration}%
  \BibitemOpen
  \bibfield  {author} {\bibinfo {author} {\bibfnamefont {T.}~\bibnamefont
  {Yamamoto}}, \bibinfo {author} {\bibfnamefont {Y.~A.}\ \bibnamefont
  {Pashkin}}, \bibinfo {author} {\bibfnamefont {O.}~\bibnamefont {Astafiev}},
  \bibinfo {author} {\bibfnamefont {Y.}~\bibnamefont {Nakamura}}, \ and\
  \bibinfo {author} {\bibfnamefont {J.-S.}\ \bibnamefont {Tsai}},\ }\href
  {\doibase 10.1038/nature02015} {\bibfield  {journal} {\bibinfo  {journal}
  {Nature}\ }\textbf {\bibinfo {volume} {425}},\ \bibinfo {pages} {941}
  (\bibinfo {year} {2003})}\BibitemShut {NoStop}%
\bibitem [{\citenamefont {Tipsmark}\ \emph {et~al.}(2011)\citenamefont
  {Tipsmark}, \citenamefont {Dong}, \citenamefont {Laghaout}, \citenamefont
  {Marek}, \citenamefont {Je\ifmmode~\check{z}\else \v{z}\fi{}ek},\ and\
  \citenamefont {Andersen}}]{PhysRevA.84.050301}%
  \BibitemOpen
  \bibfield  {author} {\bibinfo {author} {\bibfnamefont {A.}~\bibnamefont
  {Tipsmark}}, \bibinfo {author} {\bibfnamefont {R.}~\bibnamefont {Dong}},
  \bibinfo {author} {\bibfnamefont {A.}~\bibnamefont {Laghaout}}, \bibinfo
  {author} {\bibfnamefont {P.}~\bibnamefont {Marek}}, \bibinfo {author}
  {\bibfnamefont {M.}~\bibnamefont {Je\ifmmode~\check{z}\else \v{z}\fi{}ek}}, \
  and\ \bibinfo {author} {\bibfnamefont {U.~L.}\ \bibnamefont {Andersen}},\
  }\href {\doibase 10.1103/PhysRevA.84.050301} {\bibfield  {journal} {\bibinfo
  {journal} {Phys. Rev. A}\ }\textbf {\bibinfo {volume} {84}},\ \bibinfo
  {pages} {050301} (\bibinfo {year} {2011})}\BibitemShut {NoStop}%
\bibitem [{\citenamefont {Podoshvedov}(2013)}]{PhysRevA.87.012307}%
  \BibitemOpen
  \bibfield  {author} {\bibinfo {author} {\bibfnamefont {S.~A.}\ \bibnamefont
  {Podoshvedov}},\ }\href {\doibase 10.1103/PhysRevA.87.012307} {\bibfield
  {journal} {\bibinfo  {journal} {Phys. Rev. A}\ }\textbf {\bibinfo {volume}
  {87}},\ \bibinfo {pages} {012307} (\bibinfo {year} {2013})}\BibitemShut
  {NoStop}%
\bibitem [{\citenamefont {Larsen}\ \emph {et~al.}(2021)\citenamefont {Larsen},
  \citenamefont {Guo}, \citenamefont {Breum}, \citenamefont
  {Neergaard-Nielsen},\ and\ \citenamefont {Andersen}}]{Larsen}%
  \BibitemOpen
  \bibfield  {author} {\bibinfo {author} {\bibfnamefont {M.~V.}\ \bibnamefont
  {Larsen}}, \bibinfo {author} {\bibfnamefont {X.}~\bibnamefont {Guo}},
  \bibinfo {author} {\bibfnamefont {C.~R.}\ \bibnamefont {Breum}}, \bibinfo
  {author} {\bibfnamefont {J.~S.}\ \bibnamefont {Neergaard-Nielsen}}, \ and\
  \bibinfo {author} {\bibfnamefont {U.~L.}\ \bibnamefont {Andersen}},\ }\href
  {\doibase 10.1038/s41567-021-01296-y} {\bibfield  {journal} {\bibinfo
  {journal} {Nature Physics}\ }\textbf {\bibinfo {volume} {17}},\ \bibinfo
  {pages} {1018} (\bibinfo {year} {2021})}\BibitemShut {NoStop}%
\bibitem [{\citenamefont {Rauschenbeutel}\ \emph {et~al.}(1999)\citenamefont
  {Rauschenbeutel}, \citenamefont {Nogues}, \citenamefont {Osnaghi},
  \citenamefont {Bertet}, \citenamefont {Brune}, \citenamefont {Raimond},\ and\
  \citenamefont {Haroche}}]{PhysRevLett.83.5166}%
  \BibitemOpen
  \bibfield  {author} {\bibinfo {author} {\bibfnamefont {A.}~\bibnamefont
  {Rauschenbeutel}}, \bibinfo {author} {\bibfnamefont {G.}~\bibnamefont
  {Nogues}}, \bibinfo {author} {\bibfnamefont {S.}~\bibnamefont {Osnaghi}},
  \bibinfo {author} {\bibfnamefont {P.}~\bibnamefont {Bertet}}, \bibinfo
  {author} {\bibfnamefont {M.}~\bibnamefont {Brune}}, \bibinfo {author}
  {\bibfnamefont {J.~M.}\ \bibnamefont {Raimond}}, \ and\ \bibinfo {author}
  {\bibfnamefont {S.}~\bibnamefont {Haroche}},\ }\href {\doibase
  10.1103/PhysRevLett.83.5166} {\bibfield  {journal} {\bibinfo  {journal}
  {Phys. Rev. Lett.}\ }\textbf {\bibinfo {volume} {83}},\ \bibinfo {pages}
  {5166} (\bibinfo {year} {1999})}\BibitemShut {NoStop}%
\bibitem [{\citenamefont {Lemr}\ \emph {et~al.}(2011)\citenamefont {Lemr},
  \citenamefont {\ifmmode~\check{C}\else \v{C}\fi{}ernoch}, \citenamefont
  {Soubusta}, \citenamefont {Kieling}, \citenamefont {Eisert},\ and\
  \citenamefont {Du\ifmmode~\check{s}\else
  \v{s}\fi{}ek}}]{PhysRevLett.106.013602}%
  \BibitemOpen
  \bibfield  {author} {\bibinfo {author} {\bibfnamefont {K.}~\bibnamefont
  {Lemr}}, \bibinfo {author} {\bibfnamefont {A.}~\bibnamefont
  {\ifmmode~\check{C}\else \v{C}\fi{}ernoch}}, \bibinfo {author} {\bibfnamefont
  {J.}~\bibnamefont {Soubusta}}, \bibinfo {author} {\bibfnamefont
  {K.}~\bibnamefont {Kieling}}, \bibinfo {author} {\bibfnamefont
  {J.}~\bibnamefont {Eisert}}, \ and\ \bibinfo {author} {\bibfnamefont
  {M.}~\bibnamefont {Du\ifmmode~\check{s}\else \v{s}\fi{}ek}},\ }\href
  {\doibase 10.1103/PhysRevLett.106.013602} {\bibfield  {journal} {\bibinfo
  {journal} {Phys. Rev. Lett.}\ }\textbf {\bibinfo {volume} {106}},\ \bibinfo
  {pages} {013602} (\bibinfo {year} {2011})}\BibitemShut {NoStop}%
\bibitem [{\citenamefont {Heuck}\ \emph {et~al.}(2020)\citenamefont {Heuck},
  \citenamefont {Jacobs},\ and\ \citenamefont
  {Englund}}]{PhysRevLett.124.160501}%
  \BibitemOpen
  \bibfield  {author} {\bibinfo {author} {\bibfnamefont {M.}~\bibnamefont
  {Heuck}}, \bibinfo {author} {\bibfnamefont {K.}~\bibnamefont {Jacobs}}, \
  and\ \bibinfo {author} {\bibfnamefont {D.~R.}\ \bibnamefont {Englund}},\
  }\href {\doibase 10.1103/PhysRevLett.124.160501} {\bibfield  {journal}
  {\bibinfo  {journal} {Phys. Rev. Lett.}\ }\textbf {\bibinfo {volume} {124}},\
  \bibinfo {pages} {160501} (\bibinfo {year} {2020})}\BibitemShut {NoStop}%
\bibitem [{\citenamefont {Leonhardt}(1995)}]{PhysRevLett.74.4101}%
  \BibitemOpen
  \bibfield  {author} {\bibinfo {author} {\bibfnamefont {U.}~\bibnamefont
  {Leonhardt}},\ }\href {\doibase 10.1103/PhysRevLett.74.4101} {\bibfield
  {journal} {\bibinfo  {journal} {Phys. Rev. Lett.}\ }\textbf {\bibinfo
  {volume} {74}},\ \bibinfo {pages} {4101} (\bibinfo {year}
  {1995})}\BibitemShut {NoStop}%
\bibitem [{\citenamefont {Gross}\ \emph {et~al.}(2010)\citenamefont {Gross},
  \citenamefont {Liu}, \citenamefont {Flammia}, \citenamefont {Becker},\ and\
  \citenamefont {Eisert}}]{PhysRevLett.105.150401}%
  \BibitemOpen
  \bibfield  {author} {\bibinfo {author} {\bibfnamefont {D.}~\bibnamefont
  {Gross}}, \bibinfo {author} {\bibfnamefont {Y.-K.}\ \bibnamefont {Liu}},
  \bibinfo {author} {\bibfnamefont {S.~T.}\ \bibnamefont {Flammia}}, \bibinfo
  {author} {\bibfnamefont {S.}~\bibnamefont {Becker}}, \ and\ \bibinfo {author}
  {\bibfnamefont {J.}~\bibnamefont {Eisert}},\ }\href {\doibase
  10.1103/PhysRevLett.105.150401} {\bibfield  {journal} {\bibinfo  {journal}
  {Phys. Rev. Lett.}\ }\textbf {\bibinfo {volume} {105}},\ \bibinfo {pages}
  {150401} (\bibinfo {year} {2010})}\BibitemShut {NoStop}%
\bibitem [{\citenamefont {Rambach}\ \emph {et~al.}(2021)\citenamefont
  {Rambach}, \citenamefont {Qaryan}, \citenamefont {Kewming}, \citenamefont
  {Ferrie}, \citenamefont {White},\ and\ \citenamefont
  {Romero}}]{PhysRevLett.126.100402}%
  \BibitemOpen
  \bibfield  {author} {\bibinfo {author} {\bibfnamefont {M.}~\bibnamefont
  {Rambach}}, \bibinfo {author} {\bibfnamefont {M.}~\bibnamefont {Qaryan}},
  \bibinfo {author} {\bibfnamefont {M.}~\bibnamefont {Kewming}}, \bibinfo
  {author} {\bibfnamefont {C.}~\bibnamefont {Ferrie}}, \bibinfo {author}
  {\bibfnamefont {A.~G.}\ \bibnamefont {White}}, \ and\ \bibinfo {author}
  {\bibfnamefont {J.}~\bibnamefont {Romero}},\ }\href {\doibase
  10.1103/PhysRevLett.126.100402} {\bibfield  {journal} {\bibinfo  {journal}
  {Phys. Rev. Lett.}\ }\textbf {\bibinfo {volume} {126}},\ \bibinfo {pages}
  {100402} (\bibinfo {year} {2021})}\BibitemShut {NoStop}%
\bibitem [{\citenamefont {Roa}\ and\ \citenamefont
  {Groiseau}(2015)}]{PhysRevA.91.012344}%
  \BibitemOpen
  \bibfield  {author} {\bibinfo {author} {\bibfnamefont {L.}~\bibnamefont
  {Roa}}\ and\ \bibinfo {author} {\bibfnamefont {C.}~\bibnamefont {Groiseau}},\
  }\href {\doibase 10.1103/PhysRevA.91.012344} {\bibfield  {journal} {\bibinfo
  {journal} {Phys. Rev. A}\ }\textbf {\bibinfo {volume} {91}},\ \bibinfo
  {pages} {012344} (\bibinfo {year} {2015})}\BibitemShut {NoStop}%
\bibitem [{\citenamefont {Yu}\ \emph {et~al.}(2006)\citenamefont {Yu},
  \citenamefont {Song},\ and\ \citenamefont {Wang}}]{PhysRevA.73.022340}%
  \BibitemOpen
  \bibfield  {author} {\bibinfo {author} {\bibfnamefont {C.-s.}\ \bibnamefont
  {Yu}}, \bibinfo {author} {\bibfnamefont {H.-s.}\ \bibnamefont {Song}}, \ and\
  \bibinfo {author} {\bibfnamefont {Y.-h.}\ \bibnamefont {Wang}},\ }\href
  {\doibase 10.1103/PhysRevA.73.022340} {\bibfield  {journal} {\bibinfo
  {journal} {Phys. Rev. A}\ }\textbf {\bibinfo {volume} {73}},\ \bibinfo
  {pages} {022340} (\bibinfo {year} {2006})}\BibitemShut {NoStop}%
\bibitem [{\citenamefont {DiVincenzo}(1995)}]{PhysRevA.51.1015}%
  \BibitemOpen
  \bibfield  {author} {\bibinfo {author} {\bibfnamefont {D.~P.}\ \bibnamefont
  {DiVincenzo}},\ }\href {\doibase 10.1103/PhysRevA.51.1015} {\bibfield
  {journal} {\bibinfo  {journal} {Phys. Rev. A}\ }\textbf {\bibinfo {volume}
  {51}},\ \bibinfo {pages} {1015} (\bibinfo {year} {1995})}\BibitemShut
  {NoStop}%
\bibitem [{\citenamefont {Sleator}\ and\ \citenamefont
  {Weinfurter}(1995)}]{PhysRevLett.74.4087}%
  \BibitemOpen
  \bibfield  {author} {\bibinfo {author} {\bibfnamefont {T.}~\bibnamefont
  {Sleator}}\ and\ \bibinfo {author} {\bibfnamefont {H.}~\bibnamefont
  {Weinfurter}},\ }\href {\doibase 10.1103/PhysRevLett.74.4087} {\bibfield
  {journal} {\bibinfo  {journal} {Phys. Rev. Lett.}\ }\textbf {\bibinfo
  {volume} {74}},\ \bibinfo {pages} {4087} (\bibinfo {year}
  {1995})}\BibitemShut {NoStop}%
\bibitem [{\citenamefont {Barenco}\ \emph {et~al.}(1995)\citenamefont
  {Barenco}, \citenamefont {Bennett}, \citenamefont {Cleve}, \citenamefont
  {DiVincenzo}, \citenamefont {Margolus}, \citenamefont {Shor}, \citenamefont
  {Sleator}, \citenamefont {Smolin},\ and\ \citenamefont
  {Weinfurter}}]{PhysRevA.52.3457}%
  \BibitemOpen
  \bibfield  {author} {\bibinfo {author} {\bibfnamefont {A.}~\bibnamefont
  {Barenco}}, \bibinfo {author} {\bibfnamefont {C.~H.}\ \bibnamefont
  {Bennett}}, \bibinfo {author} {\bibfnamefont {R.}~\bibnamefont {Cleve}},
  \bibinfo {author} {\bibfnamefont {D.~P.}\ \bibnamefont {DiVincenzo}},
  \bibinfo {author} {\bibfnamefont {N.}~\bibnamefont {Margolus}}, \bibinfo
  {author} {\bibfnamefont {P.}~\bibnamefont {Shor}}, \bibinfo {author}
  {\bibfnamefont {T.}~\bibnamefont {Sleator}}, \bibinfo {author} {\bibfnamefont
  {J.~A.}\ \bibnamefont {Smolin}}, \ and\ \bibinfo {author} {\bibfnamefont
  {H.}~\bibnamefont {Weinfurter}},\ }\href {\doibase 10.1103/PhysRevA.52.3457}
  {\bibfield  {journal} {\bibinfo  {journal} {Phys. Rev. A}\ }\textbf {\bibinfo
  {volume} {52}},\ \bibinfo {pages} {3457} (\bibinfo {year}
  {1995})}\BibitemShut {NoStop}%
\end{thebibliography}%
\end{document}